\documentclass[aps,prd,preprintnumbers,twocolumn,eqsecnum,nofootinbib,a4paper,superscriptaddress]{revtex4-1}

\usepackage{color}
\usepackage{calc}
\usepackage{amsmath,amssymb,graphicx}
\usepackage{amssymb,amsmath}
\usepackage{tensor}
\usepackage{enumitem}
\usepackage{bm}
\usepackage{xtab}
\usepackage{float}
\usepackage{anyfontsize}
\usepackage{multirow}
\usepackage{afterpage}
\usepackage{microtype}
\usepackage{booktabs}
\usepackage{makecell}
\usepackage{multirow}
\usepackage{times}
\usepackage[varg]{txfonts}
\usepackage{musicography}
\usepackage[colorlinks, pdfborder={0 0 0}]{hyperref}
\usepackage{subfigure}
\definecolor{LinkColor}{rgb}{0.75, 0, 0}
\definecolor{CiteColor}{rgb}{0, 0.5, 0.5}
\definecolor{UrlColor}{rgb}{0, 0, 0.75}
\hypersetup{linkcolor=LinkColor}
\hypersetup{citecolor=CiteColor}
\hypersetup{urlcolor=UrlColor}
\allowdisplaybreaks
\usepackage[utf8]{inputenc}
\usepackage{ulem}
\usepackage{longtable}

\usepackage{comment}
\usepackage[nolist]{acronym}
\usepackage{tikz}
\usepackage{textcomp}
\usepackage[export]{adjustbox}
\usepackage{url}

\newcommand{\Lambdal}{\Lambda_\ell}
\newcommand{\Dl}{D_\ell}

\newcommand{\zl}{z_\ell}
\newcommand{\zs}{z_s}
\newcommand{\Ds}{D_s}
\newcommand{\Dls}{D_{\ell s}}
\newcommand{\vOmega}{\vec{\Omega}}
\newcommand{\vtheta}{\vec{\theta}}
\newcommand{\vthetal}{\vec{\theta}_\ell}

\newcolumntype{M}[1]{>{\centering\arraybackslash}m{#1}}
\newcolumntype{N}{@{}m{0pt}@{}}

\usetikzlibrary{arrows,shapes,trees,decorations.pathreplacing}
\usepackage{aas_macros}

\allowdisplaybreaks

\begin{document}	
	
\title{Strong lensing cosmography using binary-black-hole mergers: Prospects for the near future}

\author{Koustav N. Maity}
\affiliation{International Centre for Theoretical Science, Tata Institute of Fundamental Research, Bangalore 560089, India}

\author{Souvik Jana}
\affiliation{International Centre for Theoretical Science, Tata Institute of Fundamental Research, Bangalore 560089, India}
\affiliation{Department of Physics, The Chinese University of Hong Kong, Shatin, NT, Hong Kong}

\author{Tejaswi Venumadhav}
\affiliation{Department of Physics, University of California at Santa Barbara, Santa Barbara, CA 93106, USA}
\affiliation{International Centre for Theoretical Science, Tata Institute of Fundamental Research, Bangalore 560089, India}

\author{Ankur Barsode}
\affiliation{International Centre for Theoretical Science, Tata Institute of Fundamental Research, Bangalore 560089, India}

\author{Parameswaran Ajith}
\affiliation{International Centre for Theoretical Science, Tata Institute of Fundamental Research, Bangalore 560089, India}

\begin{abstract}
A small fraction of gravitational-wave (GW) signals from binary black holes (BBHs) will be gravitationally lensed by intervening galaxies and galaxy clusters. Strong lensing will produce multiple identical copies of the GW signal arriving at different times. Jana et al.~\cite{Jana_2023} recently proposed a method to constrain cosmological parameters using strongly lensed GW events detected by next-generation (XG) detectors. The idea is that the number of strongly lensed GW events and the distribution of their lensing time delays encode imprints of the cosmological parameters. From the observed number of lensed GW events (tens of thousands) and their time delay distribution, this method can provide a new probe of cosmology, obtaining information at intermediate redshifts.  In this work, we explore the possibility of doing lensing cosmography using upcoming observations of the upgraded LIGO-Virgo-KAGRA (LVK) network. This requires incorporating the detector network selection effects in the analysis, which was neglected earlier. We expect dozens of lensed GW events to be detected by upgraded LVK detectors, potentially enabling modest constraints on cosmological parameters. Even with relatively modest numbers of lensed detections, we demonstrate the potential of lensing cosmography. For XG detectors, our revised forecasts are consistent with the earlier forecasts that neglected the selection effects.
\end{abstract}

\maketitle

\section{Introduction}\label{sec:introduction}

Over the past decade, LIGO~\cite{aasi2015advanced}  and Virgo~\cite{acernese2014advanced} have detected over $200$ gravitational-wave (GW) signals from merging binaries of black holes and neutron stars~\cite{abbott2019gwtc, abbott2021gwtc, abbott2024gwtc, ligo2023gwtc, abac2025gwtc, venumadhav2020new, zackay2021detecting, olsen2022new, mehta2025new, wadekar2023new, nitz20191, nitz20202, nitz20213, nitz20234, koloniari2025new}, establishing GW observations as a new tool for astronomy. Planned sensitivity improvements to existing detectors~\cite{H1L1V1-psd-O3O4O5, HLA-psd-O6, HLA-voyager}, the addition of LIGO-India to the global network at the beginning of the next decade~\cite{ligo_india_2013a, ligo_india_2022a}, and proposed next-generation (XG) observatories~\cite{Reitze_2019, evans2021horizonstudycosmicexplorer, maggiore2020science, hild2011sensitivity} will dramatically expand our observational reach. In particular, the number of detections of GW transients is expected to reach millions in the XG era~\cite{evans2021horizonstudycosmicexplorer, maggiore2020science, chen2024cosmography}, opening unprecedented opportunities for statistical studies.

This anticipated abundance of GW observations comes at a pivotal moment for cosmology. As the precision of cosmological observations has significantly improved, tensions between different measurements have sharpened rather than resolved~\cite{efstathiou2024, peebles2025status}. The most prominent is the ``{Hubble tension}'' --- the $>5\sigma$ discrepancy in the measurement of the Hubble constant $H_0$ using Type Ia supernovae~\cite{Riess_2022} and the cosmic microwave background (CMB)~\cite{planck18_A&A}. Similar, though less severe, tensions have emerged in measurements of the matter clustering amplitude $\sigma_{8}$~\cite{sigma8_tension_2025} and the matter density parameter $\Omega_{m}$. DESI's recent measurements of the baryon acoustic oscillations (BAO), along with CMB and supernova data, hint at a time-evolving dark energy (see~\cite{Di_Valentino_2025} for an excellent summary). Whether these signal systematic errors in our measurements, unknown astrophysical complications, or genuine departures from the $\Lambda$CDM model remains an open question. In the next decade or two, GW standard siren measurements could also play an important role in addressing these questions~(see, e.g., \cite{Chen:2024gdn}).

Gravitational lensing of GWs has the potential to become yet another powerful cosmological probe. A small fraction (0.1 -- 1\%) of the GW signals observable by ground-based detectors will be strongly lensed by intervening galaxies and clusters, producing multiple copies (lensed images) of the signals that arrive at different times~\cite{abbott2020prospects, ng2018precise, oguri2018effect, li2018gravitational, smith2017strong, dai2017effect, mukherjee2021impact}. The absence of lensing signatures~\cite{hannuksela2019search, LIGOScientific:2021izm, abbott2023search, ligo_scientific_collaboration_and_virgo_2024_10841987, 1992schneider, li2023targeted, mcisaac2020search, dai2020search, janquart2023follow} in current observational data is consistent with the expected lensing rates. The first detection of strongly lensed GWs is expected in the next few years~\cite{barsode2025lensing} and will become regular as detector sensitivities keep increasing, reaching $\sim 10^{4}$ in XG detectors~\cite{Jana_2024,  evans2021horizonstudycosmicexplorer, maggiore2020science}.
  
The exact number of lensed events as well as the distribution of the lensing time delay will depend on the cosmological parameters, apart from the population properties of the sources and lenses. Jana et al.~\cite{Jana_2023} proposed a method to measure cosmological parameters from a population of lensed binary black hole (BBH) mergers detected by XG detectors in the future, and argued that the expected constraints from XG detectors will be comparable to other cosmological measurements, but probing a different cosmological epoch that is not commonly probed by other observations ($z \sim 1$--$10$).

Jana et al.~\cite{Jana_2023} assumed an {idealized} XG detector network, assuming all BBH mergers will be detectable. Since the arrival times of GW transients are measured with millisecond precision, time delays between the lensed images will be known with practically no measurement error. They also assumed that GW source population properties will be known from the large number of unlensed BBH mergers, and that the lens population will be modeled well using cosmological simulations aided with galaxy surveys. In a follow-up work~\cite{Jana_2024}, they investigated the effects of measurement errors in source and lens population properties, the effects of data contamination due to imperfect identification of lensed GW signals, and suggested ways of remedying them.

However, the path from current detectors to XG facilities spans more than a decade long and involves multiple intermediate upgrades with varying sensitivities, changing network configurations, and observation gaps for commissioning. During this evolution, strongly lensed events will accumulate gradually. In this work, we ask whether even modest constraints on cosmological parameters could be obtained from improved versions of current detectors. This requires a careful treatment of the GW detector selection effects, as these detectors will only detect loud GW sources from the relatively nearby universe (modest {horizon} distance). Using Bayesian inference on mock populations of strongly lensed events, we obtain the expected constraints on cosmological parameters (assuming a flat $\Lambda$CDM model) from upcoming observing runs of current-generation detectors, and ultimately from XG detectors. As we accumulate more strongly lensed pairs, these constraints systematically tighten. By the XG era, the precision approaches that obtained in current CMB observations and is comparable to the expected constraints from GW standard sirens.

Our most important results are summarized in Figs.~\ref{fig:lfraction_and_mrate} and \ref{fig:errors_competative}. Figure~\ref{fig:lfraction_and_mrate} shows a forecast of the expected number of BBH detections in various upcoming observing runs of LVK and XG detectors, as predicted by different astrophysical models calibrated to the current measurement of BBH merger rate at low-redshifts. The figure also shows our current estimates of the expected number of strongly lensed BBH mergers. Figure~\ref{fig:errors_competative} compares the constraints on the Hubble constant expected from GW lensing cosmography, along with the same from some of the other cosmological probes, including GW standard sirens. Figures~\ref{fig:all_combined_2D_posteriors} and \ref{fig:3D_posteriors} show the expected posteriors on the other cosmological parameters from GW cosmography. 

The paper is organized as follows. Section~\ref{sec:strongly_lensed_expectation} presents the expected detection rates of BBH mergers, as well as the fraction of strongly lensed mergers and the expected distribution of the lensing time delay. Section~\ref{sec:bayes_formalism} reviews the Bayesian inference framework for inferring cosmological parameters from lensing observables. Section~\ref{sec:results} presents the expected precision in the measurement of cosmological parameters from different observing runs. We conclude in Section~\ref{sec:outlook_and_conclusion} with an outlook discussing extensions, systematic uncertainties, and the future of GW lensing cosmography.

\section{BBH Merger rates and strong lensing probability}
\label{sec:strongly_lensed_expectation}

\begin{table*}[t]
	\centering
	\renewcommand{\arraystretch}{1.1}
	\begin{tabular}{cccc}
		\hline \hline
		\begin{tabular}[c]{@{}c@{}}Detector Network\end{tabular} &
		Detector locations (sensitivity) &
		References to Noise Curves &
		\begin{tabular}[c]{@{}c@{}}Observing period (years)\end{tabular} \\ \hline
		O4      & \begin{tabular}[c]{@{}c@{}}LIGO Hanford (O4a), LIGO Livingston (O4a), \\ Virgo Italy (O4)\end{tabular}
		& \cite{aasi2015advanced, acernese2014advanced, HLV-psd-O4a}      & 2 \\ \hline
		O5      & \begin{tabular}[c]{@{}c@{}}LIGO Hanford (A+),  LIGO Livingston (A+), \\ Virgo Italy (AdV+),  KAGRA Japan (KAGRA+)\end{tabular} 
		& \cite{aasi2015advanced, acernese2014advanced, H1L1V1-psd-O3O4O5} & 3 \\ \hline
		O6      & \begin{tabular}[c]{@{}c@{}}LIGO Hanford (A$^{\sharp}$), LIGO Livingston (A$^{\sharp}$),\\ LIGO India (A$^{\sharp}$)\end{tabular}
		& \cite{lalsuite_det_locs, aasi2015advanced, acernese2014advanced, HLA-psd-O6}         & 5 \\ \hline
		Voyager & \begin{tabular}[c]{@{}c@{}}LIGO Hanford (Voyager), LIGO Livingston (Voyager),\\ LIGO India (Voyager)\end{tabular}
		& \cite{lalsuite_det_locs, aasi2015advanced, acernese2014advanced, HLA-voyager}    & 5 \\ \hline
		XG &
		\begin{tabular}[c]{@{}c@{}} Einstein Telescope Italy (ET-D), \\ Cosmic Explorer fiducial US site (CE1 - 40 km, CE2 - 20 km)\end{tabular} 
		& \cite{lalsuite_det_locs, Borhanian_2021, Reitze_2019}& 10
		\\ \hline \hline 
	\end{tabular}
	\caption{Current (O4) and proposed (O5 and beyond) observing runs considered in this work. We assume LIGO-India will join the detector network during O6 with sensitivity comparable to the Advanced LIGO detectors at that time. For Cosmic Explorer, we model one detector with 20~km arm length and another with 40~km arm length. Columns 3 and 4 provide references for the sensitivity curves (also shown in Fig.~\ref{fig:asds}) and the assumed observation periods for each observing run, respectively.}
	\label{tab:networks}
\end{table*}

\begin{figure*}[t]
\centering
\includegraphics[height=2.5in]{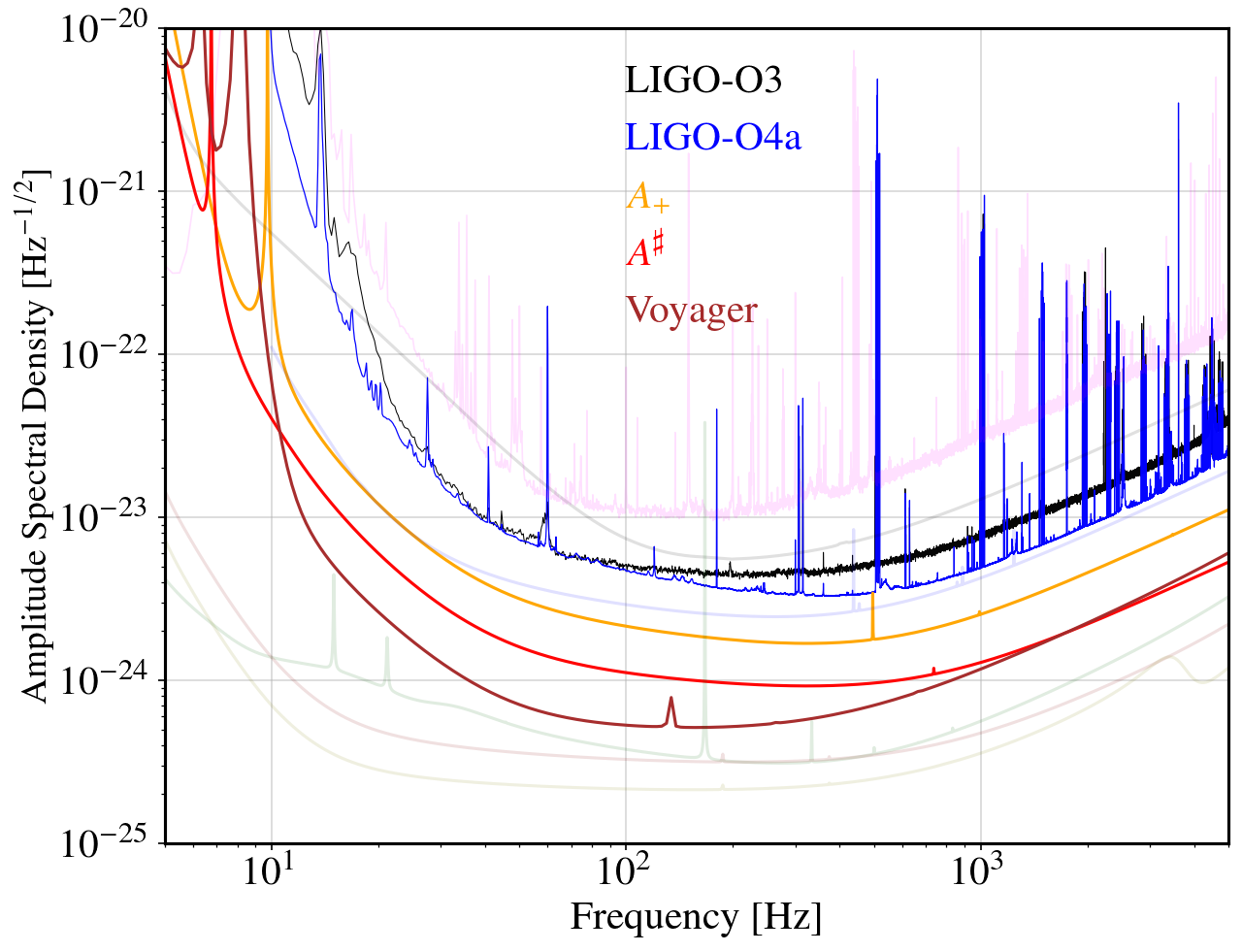}
\hspace{0.05\columnwidth}
\includegraphics[height=2.45in]{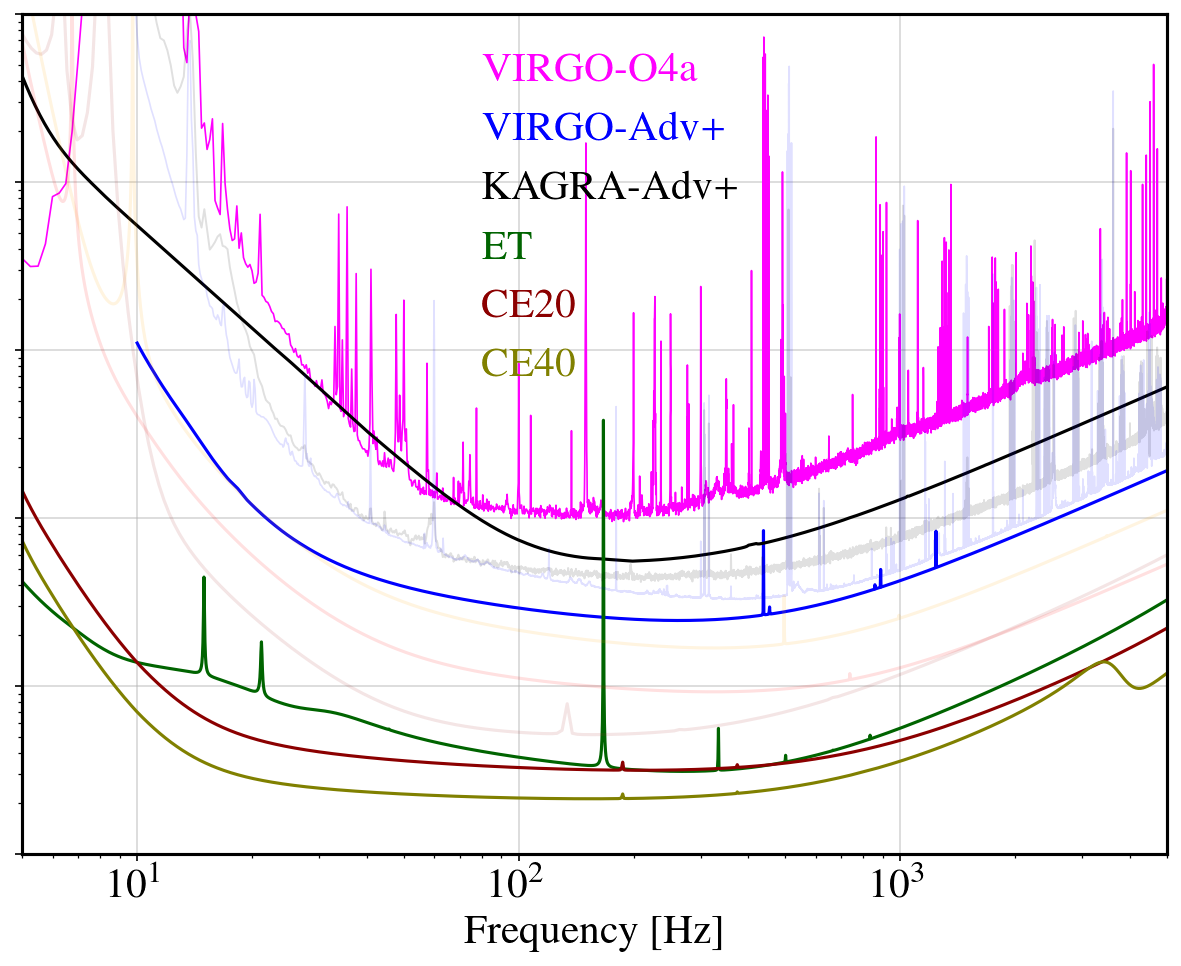}
\caption{\textit{Left:} Sensitivity curves~\cite{creighton2012} for the Advanced LIGO Livingston detector for O3, O4a, and proposed upgrades (A+, A$^{\sharp}$, and Voyager). For the Voyager configuration, we adopt the projected sensitivity curve optimized for massive binary systems~\cite{HLA-voyager}. \textit{Right:} Current and projected sensitivity curves for Virgo, along with projected sensitivity curves for KAGRA and XG detectors (CE and ET). Table~\ref{tab:networks} describes how these detectors are combined into networks with varying sensitivities across different observing runs.} 
\label{fig:asds}
\end{figure*}

\begin{figure*}[tbh]
	\centering
	\includegraphics[width=0.9\columnwidth]{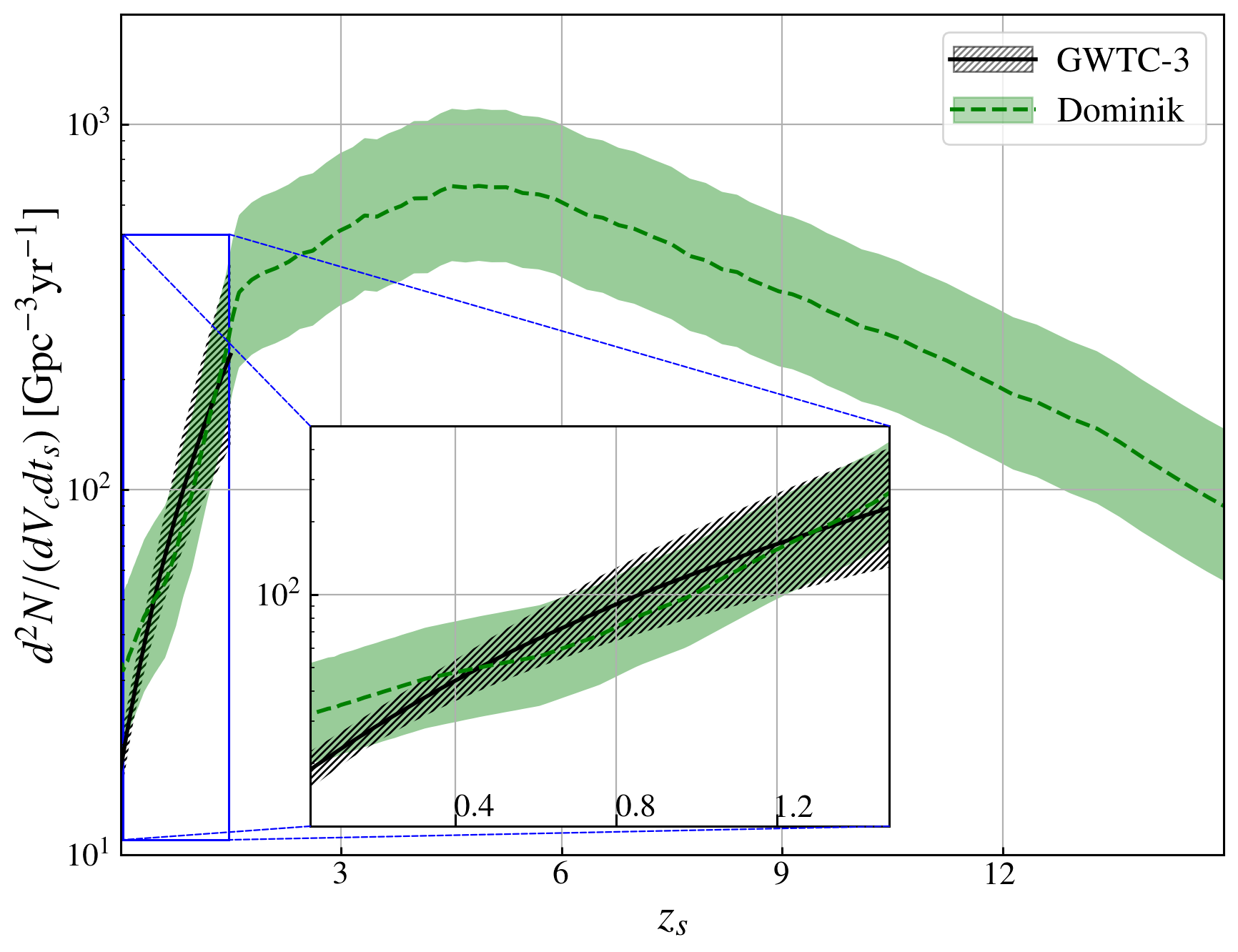}
	\hspace{0.05\columnwidth}
	\includegraphics[width=0.9\columnwidth]{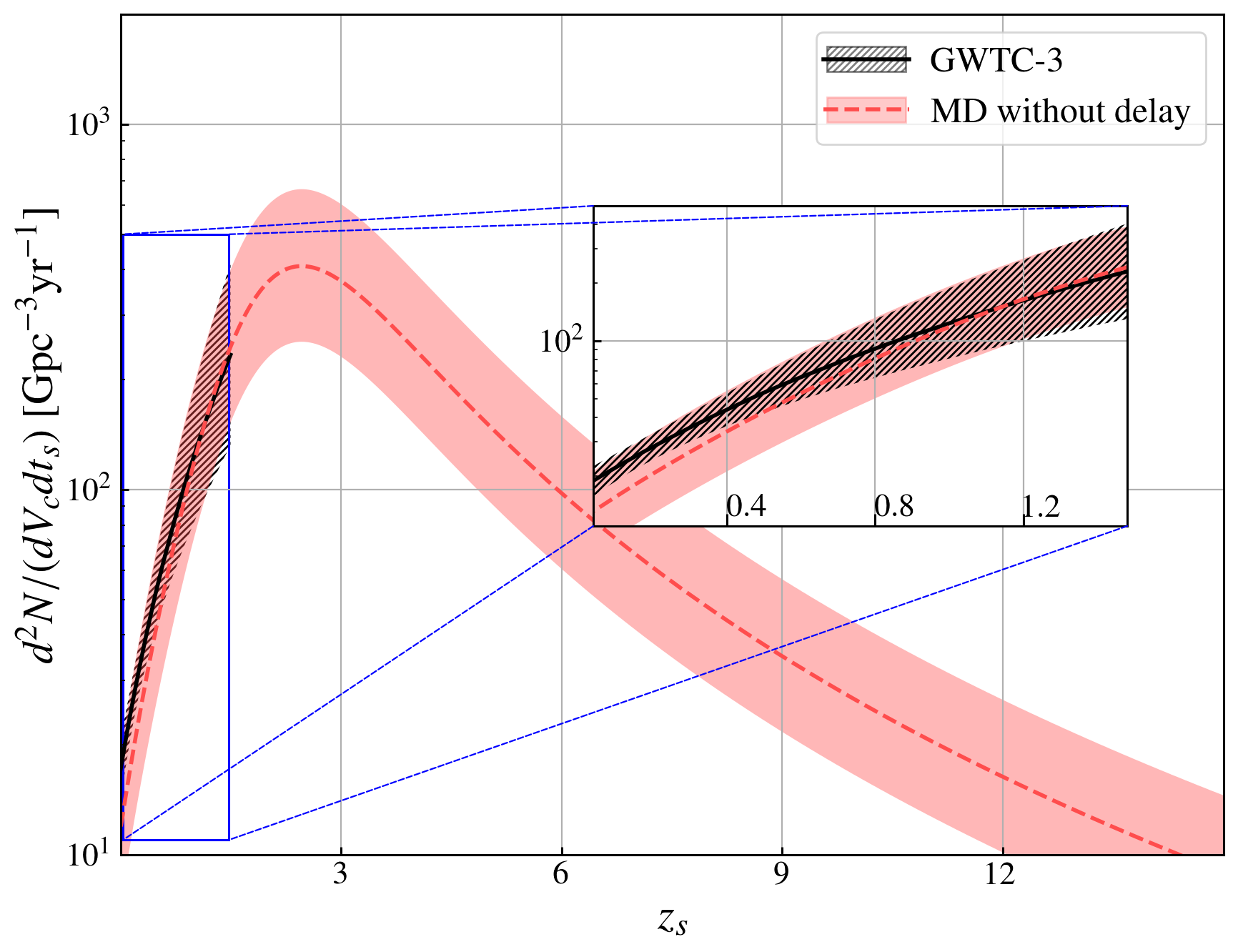}
	\caption{The source-frame merger rate density per unit comoving volume as a function of redshift, as predicted by the \emph{Dominik} model (left) and the \emph{MD without delay} model (right). Both models are normalized such that the low-redshift merger rate is consistent with the constraints from GWTC-3 data~\cite{ligo2023gwtc} (black hatched regions). Each panel includes an inset comparing the model's distribution with GWTC-3 constraints. The dashed lines correspond to the extrapolation based on the median merger rate from GWTC-3, while the shaded regions represent the central $50\%$ credible bounds. The \emph{Dominik} model exhibits relatively higher merger rates at higher redshifts, indicating a larger number of lensed GW events than the \emph{MD without delay} model.}
\label{fig:merger_rate_all_models}
\end{figure*}

In this section, we first describe the projected observation scenarios considered in this paper, including the detectors and their sensitivities. We then calculate the detectable BBH merger rates in different observing runs. Next, we introduce a formalism to compute the expected number of strongly lensed events these future networks will detect. Many quantities introduced here depend on the cosmological parameters $\vOmega$ (the Hubble constant $H_0$, the matter fraction $\Omega_m$ and the matter clustering parameter $\sigma_{8}$), assuming a flat-$\Lambda$CDM model. We exploit those dependencies to perform cosmological inferences in subsequent sections. Throughout this section, we do not explicitly show the dependence of these quantities on cosmological parameters, for brevity, and adopt the \texttt{Planck18}~\cite{planck18_A&A} cosmological parameters as our fiducial cosmology:  $H_{0} = 67.66$ km s$^{-1}$ Mpc$^{-1}$, $\Omega_{m} = 0.3111$, $\sigma_{8} = 0.8102$.

\subsection{Detector networks and sensitivities}

The fourth observing run (O4), involving Advanced LIGO and Virgo, has already detected over 200 GW candidates~\cite{GraceDB_public}. Following O4, the detector sensitivity will further improve with proposed upgrades in the fifth observing run (O5). Scheduled to commence in the late 2020s, O5 is expected to incorporate KAGRA into the detector network. Beyond O5, proposed upgrades include $A^{\sharp}$~\cite{HLA-psd-O6} and Voyager~\cite{Adhikari:2019zpy}, which aim to potentially achieve maximum sensitivity limits within existing facility constraints (cf. Fig.~\ref{fig:asds}). XG detectors like Cosmic Explorer~\cite{Reitze_2019, evans2021horizonstudycosmicexplorer} and the Einstein Telescope~\cite{maggiore2020science, hild2011sensitivity} are expected to significantly increase detection capabilities. These detectors may potentially detect millions of binaries in coming decades.

In this study, we consider different observing scenarios starting from the O4 observing run. Table~\ref{tab:networks} provides an overview of these detection scenarios, detailing the detector locations and sensitivities for each network. In Fig.~\ref{fig:asds}, we plot the corresponding expected noise amplitude spectral densities~\cite{creighton2012} of these detectors used in this work.

\subsection{BBH merger rates}

To forecast the BBH detection rates in future observing runs, we first model the intrinsic merger rate distribution per comoving volume as a function of redshift, leveraging available observational constraints at low redshifts and astrophysical models pertaining to high redshifts. Assuming perfect detectability across all distances and source parameters, this intrinsic rate represents the theoretical upper limit for detections. Second, we model the detector network's sensitivity limitations by accounting for observational selection effects. Convolving the intrinsic merger rate with the network's detection capabilities yields the GW detection rate — the expected number of events per year at given sensitivities.
    
\subsubsection{Intrinsic merger rate}

If $d^{2}N/(dV_{c}dt_{\rm s})$ is the intrinsic merger rate density per unit comoving volume in the source frame, then the detection rate in an ideal detector network\footnote{A hypothetical detector network with infinite sensitivity, would detect binary mergers throughout the entire observable universe.} can be written as
\begin{align}
	\label{eq:int_BBH_rate}
	\mathcal{R}_{\rm int} & = \dfrac{d N}{d t_{\rm d}} = \int_{0}^{z_{\rm max}}\!\!\! \dfrac{dz}{(1 + z)} ~ \dfrac{d^{2}N}{dV_{c}dt_{\rm s}}(z) ~ \dfrac{dV_{c}}{dz} (z),
\end{align}
where ${dV_{c}}/{dz}$ is the differential comoving volume and $z$ is the cosmological redshift. The factor $1/(1 + z)$ accounts for cosmological time dilation, converting from the source frame time $t_s$ to the observer frame time $t_d$.

Current GW observations provide constraints on the merger rate density at low redshifts. We use several astrophysical models to describe the merger rate evolution at higher redshifts. The overall normalizations for these models are determined via $\chi^{2}$ minimization (shown in Fig.~\ref{fig:merger_rate_all_models}), using $d^{2}N/(dV_{c}\,dt_\mathrm{s})$ inferred from the GWTC-3 data at lower redshifts ($z \leq 1.5$)~\cite{ligo2023gwtc}. In this work, we employ the following models at higher redshifts:

\begin{enumerate}[label=\alph*)]

\item \textit{Dominik}: In this model~\cite{Dominik_2013}, star formation rate follows Strolger et al.~\cite{Strolger_2004}, with stellar populations evolved using \texttt{Startrack} population synthesis code~\cite{Belczynski_2008}. \texttt{Startrack} comprehensively takes care the different astrophysical effects in a binary evolution, like mass transfer, tidal interactions, stellar winds, etc.
	
\item \textit{MD without delay}:  Given that stellar mass compact objects originates from main-sequence stars, we assume their distribution follows the Madau-Dickinson (MD) star formation rate (SFR)~\cite{Madau_2014}, peaking at $\zs \sim 2 - 3$ for standard cosmology. We also multiply an efﬁciency factor $\eta(z_{s})$, as in~\cite{Vitale_2019}, to take into account the fact that, of the stars formed at a given redshift, only a fraction with smaller metallicity will form heavier black holes. \emph{Without delay}  emphasizes the fact what we have neglected the time-delay between the black hole formation and the actual merger. 
	
\end{enumerate}  

\begin{figure}[tbh]
	\centering		
	\includegraphics[width=0.85\columnwidth]{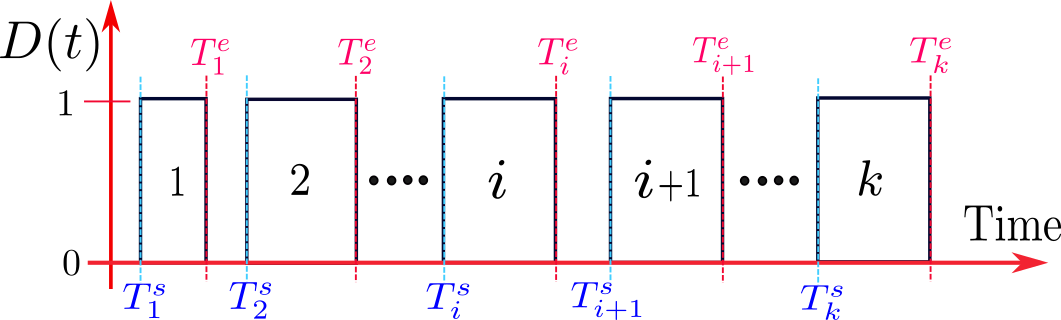}
	\caption{Duty cycle of GW detector networks. The duty cycle equals $1$ when the detectors are operational and $0$ otherwise. Each observing segment is denoted by two time-stamps: for the $i^{\rm th}$ observing run, the run begins at $T_i^{s}$ and concludes at $T_i^{e}$.}
	\label{fig:revised_duty_cycle}
\end{figure}

\begin{figure}[tbh]
\includegraphics[width=0.9\columnwidth]{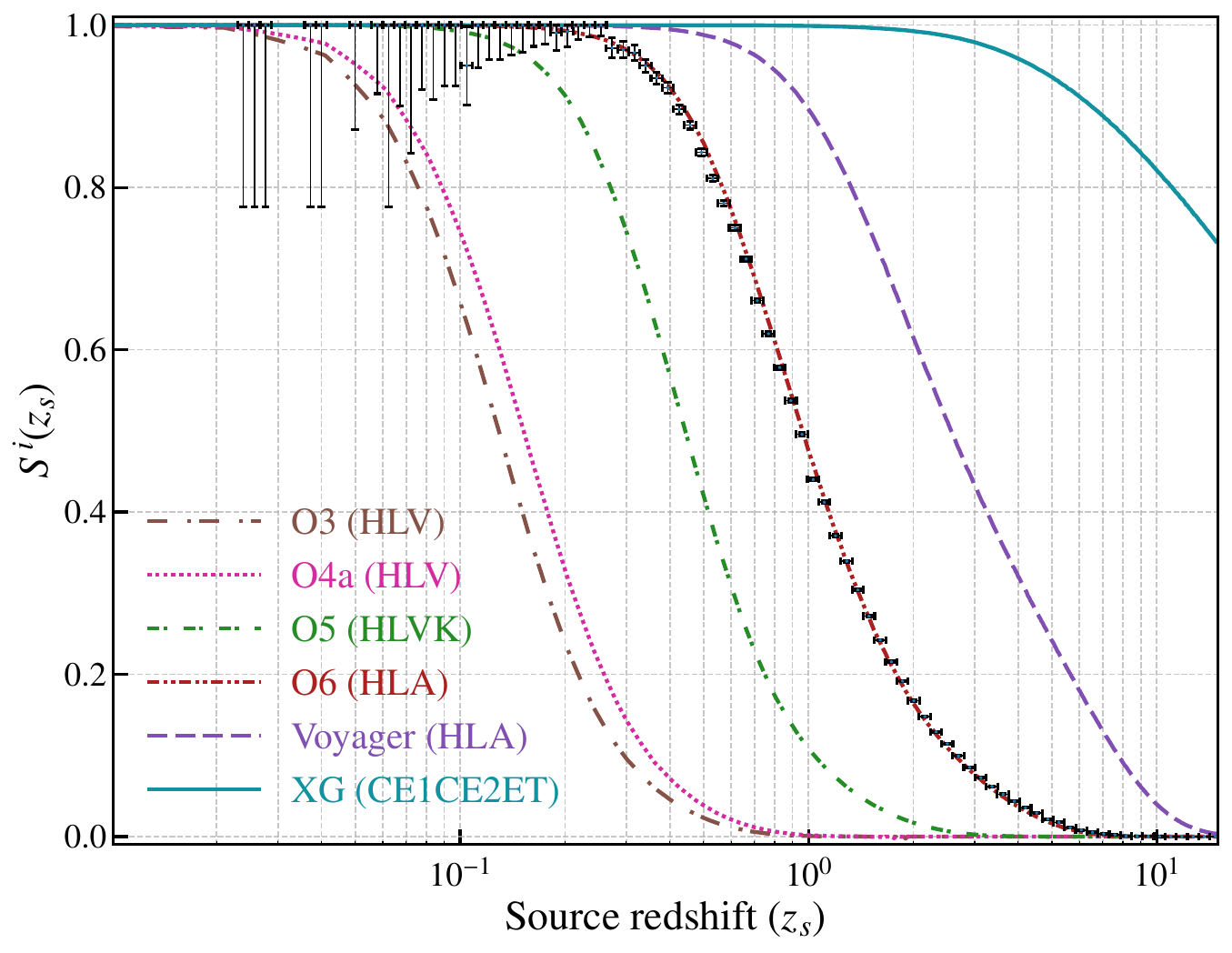}
\caption{The selection function $S^{i}(\zs)$, defined as the fraction of BBH mergers (marginalized over all other source parameters) detectable at a given source  redshift. We sample a large number ($10^{6}$) different BBH parameters at each constant redshift~(or luminosity distances, equivalently), compute the cumulative probability distribution of their optimal network SNR at the threshold. At very small redshifts, almost all events can be detected, whereas at higher redshifts the detectable fraction smoothly decreases to zero, with the cutoff redshift depending on the sensitivity of the detector network. For O6, we additionally sample the luminosity distance (following \emph{MD without delay} distribution), together with the remaining BBH parameters. We then compute the optimal network SNR for each event and compute the fraction of events exceeding the detection threshold. This explicit Monte Carlo sampling method (denoted by black scatters) agrees well with the averaged detection fraction.}
\label{fig:selection_fun}
\end{figure}

\subsubsection{Selection effects and the detectable merger rate}

In reality, GW detectors are endowed with a limited observable space-time volume that is dependent on the source characteristics. In the leading order, the measured GW strain is given by $h(t) \propto \mathcal{M}_{z}^{5/4} Q(\theta, \phi, \psi, \iota) / d_{L}$~\cite{curtler_1993, curtler_1994}. Here, $\mathcal{M}_{z}$ is the redshifted chirp mass, $Q$ is a function of angular sky location and orientation of the source with respect to a given detector, and $d_{L}$ is the luminosity distance to the source from the observer~\cite{hogg1999distance, maggiore2020science}. Therefore, in general, we expect to detect more nearby sources than faraway sources. However, depending on various combinations of the above factors, GW signals from massive, face-on ($\iota = 0$) binaries can be detected out to a larger distance, whereas less massive, edge-on ($\iota = \pi/2$) binaries will have only a smaller observable horizon. In summary, we have a \emph{selection bias} towards detecting the louder events~\cite{Malmquist_1922, Malmquist_1925}.

On average, a detector will be less sensitive to distant sources, leading to a smaller probability of detection. We assume that a GW signal is confidently detected if it exceeds a predetermined signal-to-noise ratio (SNR) threshold $\rho_{\rm th}$. We can then calculate the probability of detection, i.e. the network \textit{selection function} $S(\zs)$, by integrating the SNR distribution $d\tilde{P}/d\rho$ from $\rho_{\rm th}$.
\begin{equation}
S(\zs) = \int_{\rho_\mathrm{th}}^\infty d\rho   \frac{d\tilde{P}}{d\rho}(\zs), 
\label{eq:selection_fn_unlens}
\end{equation}
where ${d\tilde{P}}/{d\rho}(\zs)$ is marginalized over all other source parameters $\vtheta$, except $\zs$\footnote{When we include redshift as one of the extrinsic parameters of a BBH merger, a particular cosmology is implicitly assumed to transform the luminosity distance to redshift. Here, we assume the $\Lambda$CDM model with parameters given by \texttt{Planck18}~\cite{planck18_A&A} as our fiducial cosmology.}: 
\begin{equation}
\frac{d\tilde{P}}{d\rho}(\zs) = \int_0^{T_d} dt ~ \frac{dP}{dt}  D(t) \int d\vtheta ~ \frac{dP}{d\vtheta} \frac{dP}{d\rho}(\zs, \vtheta, t). 
\label{eq:snr_dist_unlens}
\end{equation}
Here, $t$ is the time of arrival of the GW signal at the earth and $T_d$ is the \textit{total} observation time, including any gaps in observation. $D(t)$ is the duty cycle of the detector network: $D(t) = 1$ when the detectors are operational and 0 otherwise (see Fig.~\ref{fig:revised_duty_cycle}). The SNR probability distribution ${dP}/{d\rho}$ depends on both intrinsic and extrinsic source parameters $\vtheta$ (expect redshift) that determine the GW signal. It is also a function of time (modulo the sidereal day) because the detector antenna patterns change due to the Earth's rotation. The astrophysical prior of the arrival time of the GW signal can be taken to be a uniform distribution: ${dP}/{dt} = 1/T_d$. However, this will be modulated by the duty cycle $D(t)$ of the network. The prior for the properties of the GW sources, ${dP}/{d\vtheta}$, should be astrophysically motivated and observationally supported. We use the relevant median distributions of different parameters from~\cite{ligo2023gwtc} (for e.g., the \textsc{Power Law + Peak} model for primary mass distribution, power law for mass ratio, etc.). 

If the start and end time of each observing segment $i$ is given by $T^s_i$ and $T^e_i$, respectively (see Fig.~\ref{fig:revised_duty_cycle}), with $k$ such segments, Eq.~\eqref{eq:snr_dist_unlens} can be simplified as 
\begin{equation}
\frac{d\tilde{P}}{d\rho}(\zs) = \sum_{i=1}^{k} \frac{T_{i}^{\rm obs}}{T_d} \int d\vtheta ~ \frac{dP}{d\vtheta} \frac{dP_i}{d\rho}(\zs, \vtheta),
\label{eq:snr_dist_unlens_simple}
\end{equation}
where $dP_{i}/d\rho$ is the SNR distribution for the $i^{\rm th}$ observing segment, while $T_{i}^{\rm obs} = T^e_i - T^s_{i} ~ $ is the duration of the  segment. We have assumed that the SNR distribution doesn't change significantly within a single segment. 

Convolving the selection function $S(\zs)$ with the intrinsic merger rate and multiplying with the total observation time $T_d$, we obtain the expected number of detectable events:
\begin{align}\label{eq:R_det}
\Lambda =  T_d  \int_{0}^{z_{\rm max}} \dfrac{d\zs}{(1 + \zs)} ~ \dfrac{d^{2}N}{dV_{c}dt_{s}}(\zs) ~ \dfrac{dV_{c}}{d\zs}(\zs) ~ S(\zs).
\end{align}
Using Eqs.~\eqref{eq:selection_fn_unlens} and \eqref{eq:snr_dist_unlens_simple}, we can re-write the above equation in the following way to get the expected number of BBH detections $\Lambda_{i}(T_{i}^{\rm obs})$ in each observing segment, 
\begin{align}\label{eq:sum_of_ind_scenarios}
\Lambda = \sum_{i = 1}^{k} \Lambda_{i} \left(T_{i}^{\rm obs}\right),
\end{align} 
where
\begin{align}\label{eq:exp_ind_det}
	\Lambda_{i}\left(T_{i}^{\rm obs}\right) = T_{i}^{\rm obs}\int_{0}^{z_{\rm max}} \!\!\!\!\! \dfrac{d\zs}{(1 + \zs)} ~ \dfrac{d^{2}N}{dV_{c}dt_{s}}(\zs) ~ \dfrac{dV_{c}}{d\zs}(\zs) ~ S^{i}(\zs)
\end{align}     
with the selection function in each detection scenario defined as 
\begin{align}\label{eq:unlensed_selection_iobs}
	S^{i}(\zs) = \int_{\rho_{\rm th}}^{\infty} \!\!\!\!\! d\rho \int \!\!\! d\vtheta ~\frac{dP}{d\vtheta} \frac{dP_{i}}{d\rho}(\zs, \vtheta).
\end{align}
If the sensitivity of the detector networks changes in different observing segments, $S^{i}(\zs)$ changes accordingly (see Fig.~\ref{fig:selection_fun}).

\subsection{Number of detectable strongly lensed pairs}

Counting the number of detectable lensed events is more complex. For example, even if we detect the first image, the other images may arrive during different observing runs, when detector sensitivities may differ. They may also arrive when detectors are not operating, and may be harder to detect due to demagnification that is typical of later images.            

In this work, we assume that the lenses are well-described by singular isothermal spheres (SIS)~\cite{1992schneider}, thus producing only two lensed images (while this is a quite simplistic model for now, it will be improved in future work.) The number of lensed mergers for which both images are detectable can be written as:
\begin{align}\label{eq:exp_nl_master_eq}
\Lambdal  = T_d  \int_{0}^{z_{\rm max}} \!\! \dfrac{d\zs}{(1 + \zs)} ~ \dfrac{d^{2}N}{dV_{c}dt_{s}} ~ \dfrac{dV_{c}}{d\zs}(\zs) ~ P_{\ell}(\zs)~S^L(\zs).
\end{align} 
In the above equation, $P_{\ell}(\zs)$ is the strong lensing probability at redshift $\zs$, which we formulate next, and, $ S^{L}(\zs)$ is the selection function for detecting both the lensed images. We will see that this depends on the duty cycles of the detector networks. 

\subsubsection{The strong lensing probability}

The probability that a source located at a redshift $\zs$ will encounter at least one strong gravitational lens is given by 
\begin{align}\label{eq:strong_opt_depth}
P_{\ell}(\zs) = 1 - e^{-\tau(\zs)}, 
\end{align}
where $\tau(\zs)$ is the strong lensing optical depth, which is the average number (Poisson mean) of strong lenses encountered by the radiation as it propagates from the source to the observer. In terms of the differential optical depth ${d\tau}/{d\zl}$, 
\begin{align}
\tau(\zs) = \int_0^{\zs} d\zl \dfrac{d\tau}{d\zl} (\zl, \zs). 
\label{eq:opt_depth}
\end{align}
The differential optical depth depends on two key factors: the number density of lenses at each redshift, and their effective strong lensing area. For a spherically symmetric lens, the effective lensing area is
\begin{equation}\label{eq:A_eff}
A_{\rm eff} (\zl, \zs, \vthetal) = \pi~y_{r}^{2}(\vthetal)~l_{\rm 0}^{2}(\zl, \zs, \vthetal).
\end{equation} 
Here, $l_{\rm 0}$ is the characteristic length scale (e.g., the Einstein radius $r_E$ for SIS lenses), $y_{r}(\vthetal)$ represents the critical source position necessary to form multiple images, in units of $l_{\rm 0}$. The differential optical depth is therefore given by:
\begin{equation}
	\dfrac{d\tau}{d\zl} (\zl, \zs) = \int d\vthetal \dfrac{d^2n}{d\vthetal d\zl} 
	\dfrac{A_{\rm eff} (\zl, \zs, \vthetal)}
	{4\pi D_{\ell}^{2}(\zl)}
	\label{eq:diff_opt_depth}
\end{equation}
where $d^2n/(d\vthetal d\zl)$ is the number of lenses within redshift interval $(\zl, \zl + d\zl)$ and parameter range $(\vthetal , \vthetal + d\vthetal)$, while $D_{\ell} (\zl)$ is the angular diameter distance from the observer to the lens at redshift $\zl$. 

The SIS lens model is characterized by a single parameter $\sigma$ --- the velocity dispersion of the particles constituting the lens. For these lenses, $y_{r} = 1$ and $l_{0} = r_{E}(\zl, \zs)$ --- sources that lie within $r_E$ of the lens  will produce multiple images, where
\begin{equation}\label{eq:einstein_radius}
	r_E(\zl, \zs) = 4\pi \left(\dfrac{\sigma}{c}\right)^{2} 
	\dfrac{D_{\ell}(\zl) D_{\ell s}(\zl, \zs)}{D_{s}(\zs)}.
\end{equation}
Above, $\Ds$ is the angular diameter distance to the source at $\zs$ and $\Dls$ is the angular diameter distance between the source and the lens. The number of lenses can be calculated as: 
\begin{equation}
	\dfrac{d^2n}{d\sigma d\zl} (\zl) = \dfrac{d^2n}{d\sigma dV_{c}}(\zl) \dfrac{dV_{c}}{d\zl} (\zl),
	\label{eq:dn_by_dsigma_dzl}
\end{equation} 
where ${d^2n}/{d\sigma dV_{c}}$ is the comoving number density within a specific velocity-dispersion range and ${dV_{c}/dz_{\ell}}$ is the differential comoving volume introduced earlier. 

We use a halo mass function (HMF) that gives the number density ${d^2n}/{dM_{h} dV_{c}}$ of dark matter halos with different masses at various redshifts, that is obtained from cosmological simulations. We combine this with a prescription to convert the mass of the dark matter halo to the velocity dispersion of the SIS lens~\cite{Jana_2023,Jana_2024}:
\begin{align}
	\dfrac{d^2n}{d\sigma dV_{c}}(\zl) = \dfrac{d^2n}{dM_{h} dV_{c}}(\zl) ~ \dfrac{dM_{h}}{d\sigma}(\zl). 
	\label{eq:num_dens_lens}
\end{align}
We use the Behroozi model~\cite{Behroozi_2013} for the HMF with halo masses in the range $M_{h} = 10^{8}-10^{15}~M_{\odot}$. For \( M_h \) to \( \sigma \) conversion, we assume that haloes are spherically symmetric and virialized at the observation epoch. Following~\cite{Jana_2023,Jana_2024}, the mean density of a halo is taken to be the virial overdensity at the lens redshift \( z_{\ell} \); that is, \( \rho(z_{\ell}) = \Delta_{c}(z_{\ell})\, \rho_{c}(z_{\ell}) \), where \( \Delta_{c}(z_{\ell}) \) is the critical overdensity factor for virialization, and \( \rho_{c}(z_{\ell}) \) is the critical density of the universe at that epoch~\cite{bryan1998}. With the density fixed, we compute the halo radius using the relation \( M_h = {4}\pi R^3 \rho(z_{\ell})/3 \). The velocity dispersion is then given as:   
\begin{equation}\label{eq:sigma_prescription}
\sigma = \sqrt{\frac{G M_h}{R}},  ~~ \implies \frac{dM_h}{d\sigma} = \frac{3M_h}{\sigma}. 
\end{equation}
Finally, combining Eqs.~\eqref{eq:A_eff} and \eqref{eq:dn_by_dsigma_dzl}, the differential optical depth (Eq.~\eqref{eq:diff_opt_depth}) becomes 
\begin{align}\label{eq:SIS_optd}
	\dfrac{d\tau}{d\zl} (\zl, \zs)   = \int_{\sigma^\mathrm{min}}^{\sigma^\mathrm{max}} d\sigma  \dfrac{d^{2}n}{d\sigma d\zl} (\zl) \dfrac{r_E^{2}(\zl, \zs)}{4D^{2}_{\ell} (\zl)}, 
\end{align}     
where $\sigma^\mathrm{min}$ and $\sigma^\mathrm{max}$ are the $\zl$ dependent velocity dispersions corresponding to the least massive ($10^{8} M_{\odot}$) and most massive ($10^{15} M_{\odot}$) haloes that we assume. Integrating Eq.~(\ref{eq:SIS_optd}) over lens redshift gives the strong lensing optical depth for SIS lenses.
 
\subsubsection{The detector network selection function for lensed events}

\begin{figure}[tbh]
\centering
\includegraphics[width=0.78\columnwidth]{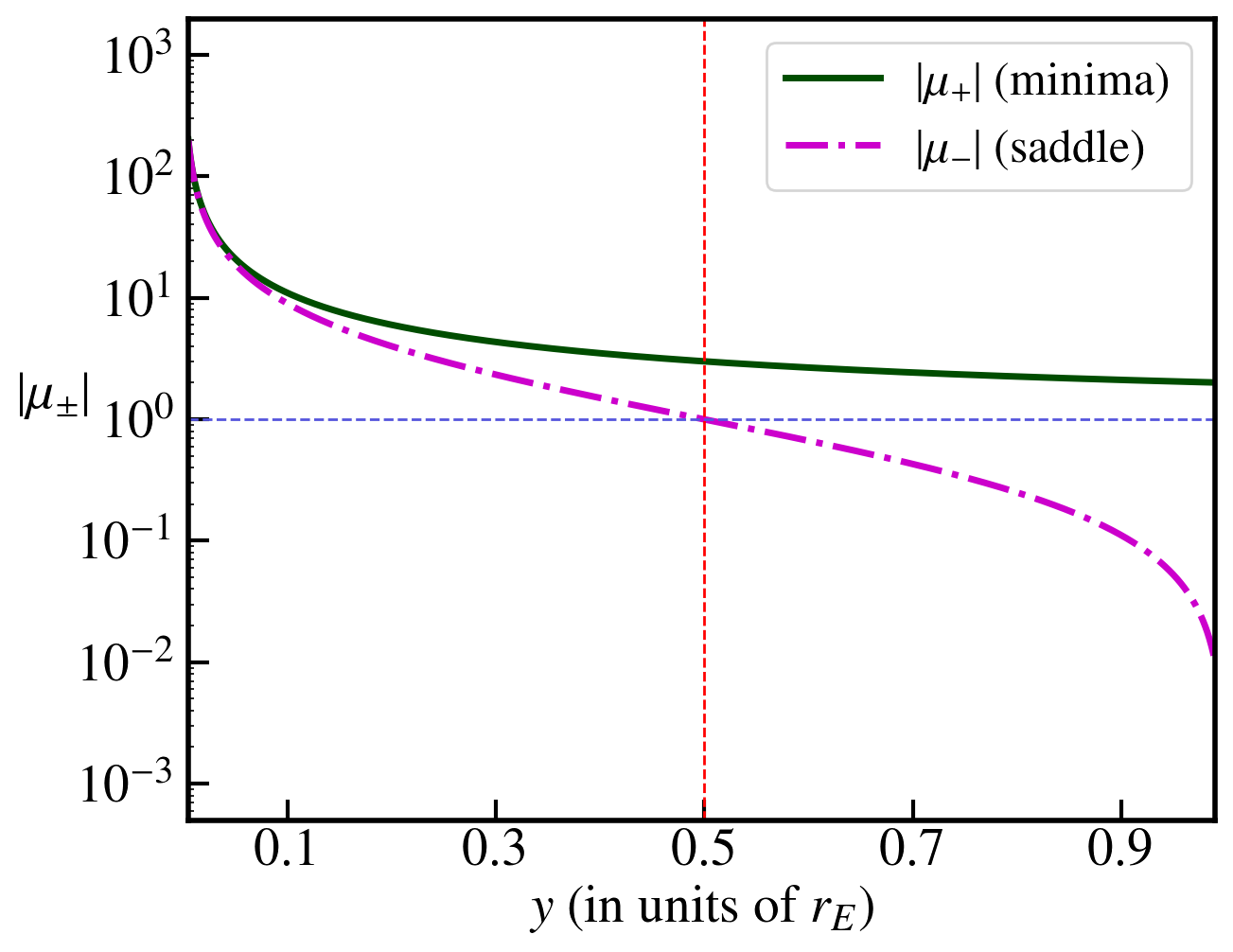}
\caption{The absolute values of the magnification of the two images produced by an SIS lens when $y < 1$, as a function of the impact parameter. The green curve corresponds to the minima of the time-delay surface, which is always detected first, while the magenta curve corresponds to the saddle. Note that both images are always magnified ($|\mu_{\pm}| > 1$) when $y < 0.5$; otherwise, only the minima images are magnified.}
\label{fig:selection_fn_inputs}
\end{figure} 

An SIS lens produces two images when the impact parameter $y < 1$ (in units of $r_E$) and one image otherwise. If an SIS lens at redshift $\zl$ with velocity dispersion $\sigma$ is responsible for strong lensing of a BBH merger at redshift $\zs$, then the time delay between the two images is given by~\cite{1992schneider}:
\begin{equation}
\label{eq:deltat-SIS}
\Delta t (\zl, \sigma, \zs, y) = \left(1+\zl \right) ~\frac{32\pi^2 y}{c} \left(\frac{\sigma}{c}\right)^4 \frac{\Dl(\zl) \, \Dls(\zl, \zs)}{\Ds(\zs)},
\end{equation}
where $\Dl$, $\Ds$, and $\Dls$ are the angular diameter distances to the lens, to the source, and between the lens and source, respectively. Magnifications of the two images are given by (Fig.~\ref{fig:selection_fn_inputs})
\begin{equation}\label{eq:mags}
\mu_+ = 1+{y}^{-1}, \qquad \mu_- = \left| 1-{y}^{-1}\right |.
\end{equation} 
If either image arrives during gaps between the observing runs, or if either has an SNR below the detection threshold, we will fail to identify the lensed pair. Additionally, since the images arrive at different times, they experience different antenna patterns due to Earth's rotation. For simplicity, we assume that if both images arrive when the detectors are operational and both exceed the SNR threshold, they are correctly identified as a lensed pair, without any false alarms. In the future, this can be easily generalized by taking into account the detection efficiency and false alarm rate of lensing searches (see, e.g.,~\cite{Barsode:2025agk}). 

In order to compute the selection function for detecting lensed events, we need to consider the joint distribution of the SNRs of the two images $d^2P/d\rho_1 d\rho_2$, which is a function of $(\zs, \vtheta, t, \Delta t)$, where $\vtheta$ is the set of source parameters describing the (unlensed) GW signal, $t$ is the time of arrival of the first image at the earth, and $\Delta t$ is the time delay between the two images. As earlier, we denote the duty cycle of the detector network by $D(t)$ (see Fig.~\ref{fig:revised_duty_cycle}).

\begin{figure}[t]
\includegraphics[width=0.85\columnwidth]{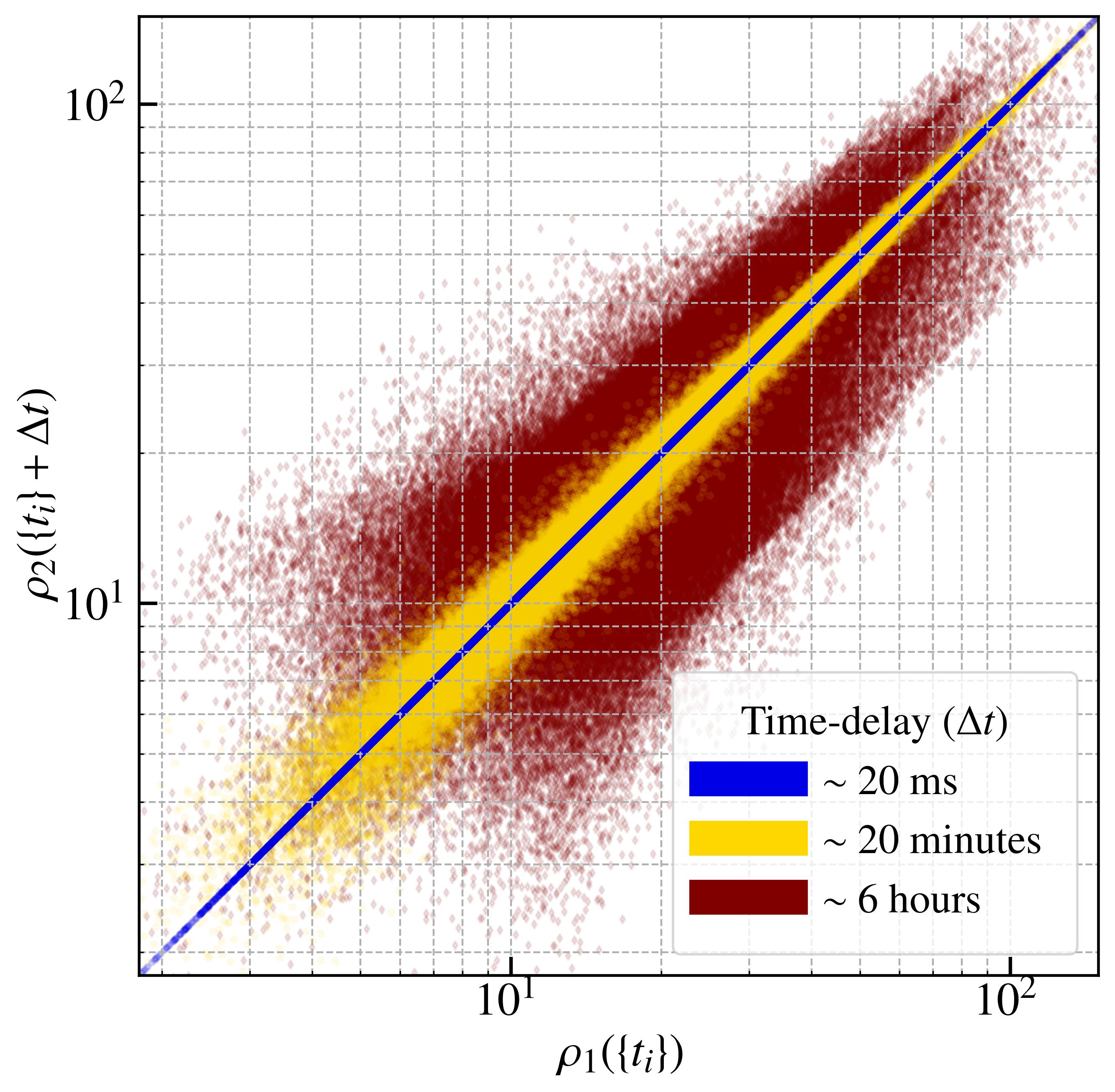}
\caption{The (unmagnified) SNR samples at a fixed redshift $z_{s} = 0.21$, assuming O5 sensitivity at three different fixed time-delays. The SNR of the first image, $\rho_{1}$, is computed with arrival times uniformly distributed across the mergers. We then add a fixed time-delay to those arrival times to compute the SNR of the second image, $\rho_{2}$.}
\label{fig:joint_unlensed_snr_2D_dist}
\end{figure}

We first compute the joint distribution of $\rho_1$ and $\rho_2$ marginalized over $\vtheta$ and $t$ 
\begin{align}
\frac{d^2\tilde{P}}{d\rho_1 d\rho_2}(\zs, \Delta t) &=  \int_0^{T_d} dt \dfrac{dP}{dt}  D(t) \, D(t+\Delta t) \nonumber \\    
& \times \int d\vtheta ~ \dfrac{dP}{d\vtheta} \dfrac{d^2P}{d\rho_1 d\rho_2} (\zs, \vtheta, t, \Delta t), 
\label{eq:snr_dist_joint}
\end{align}
where, as we did in the case of the unlensed GW events (see Eq.~\eqref{eq:snr_dist_unlens}), the prior for the arrival time of the first image can be taken to be uniform: ${dP}/{dt} = 1/T_d$. However, this will be modulated by the duty cycle of the detector networks at the time of arrival of the two images at the earth, $D(t)$ and $D(t+\Delta t)$. For the source parameters of the GW sources ${dP}/{d\vtheta}$, we use the same prior as Eq.~\eqref{eq:snr_dist_unlens}. Figure \ref{fig:joint_unlensed_snr_2D_dist} shows one realization of SNR samples for three different time-delays, assuming $\zs = 0.21$ and O5 sensitivity. We do not assume any duty-cycle modulation in the plot. For very short time-delays, the change in the antenna pattern is negligible, and both the unlensed SNRs are basically the same. As time-delay increases, these distributions start to smear out.   

\begin{figure*}[tbh]
\centering
\includegraphics[width=0.9\textwidth]{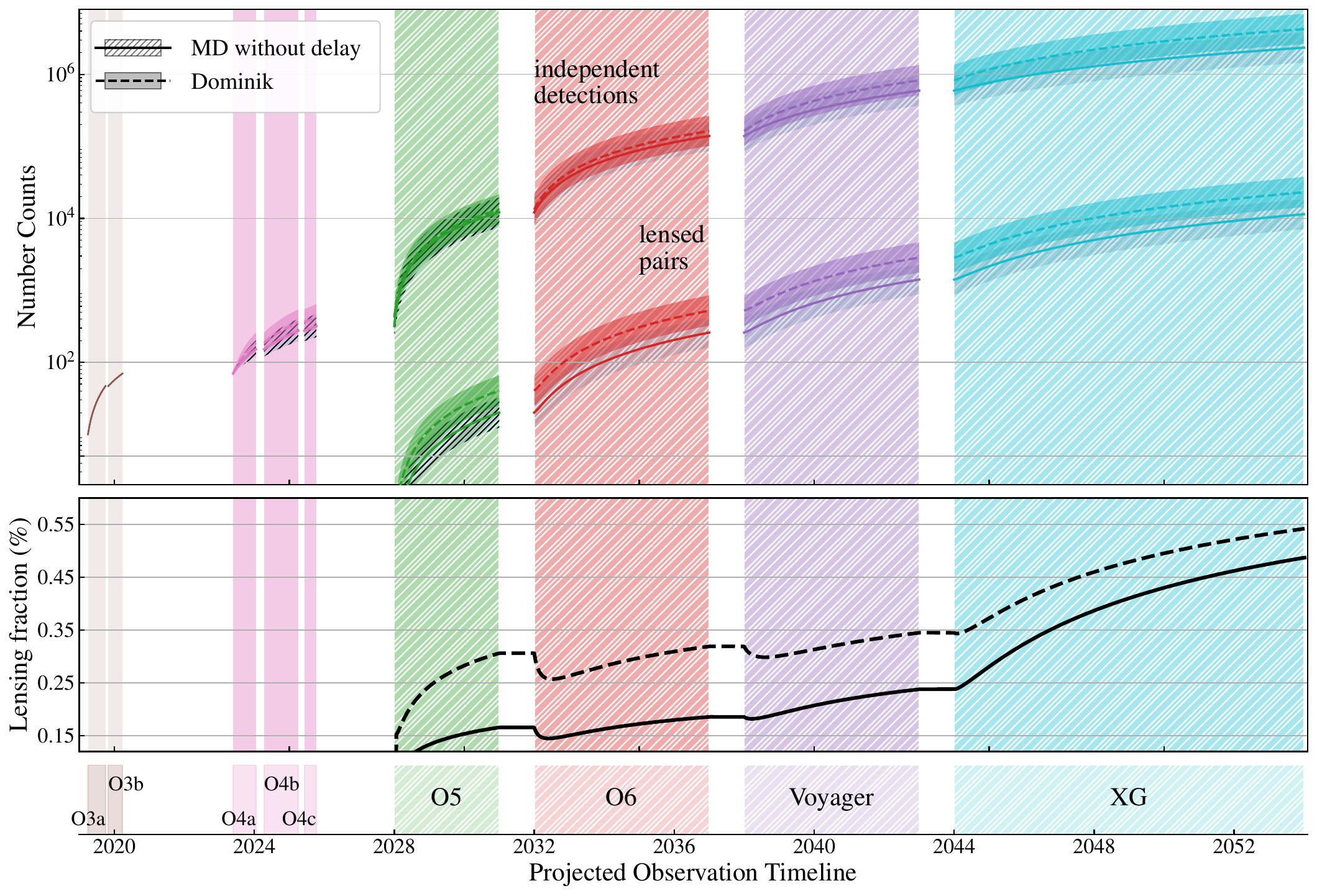}
\caption{\emph{Top panel}: Cumulative number of BBH detections (upper curves) and strongly lensed pairs (lower curves) across different observing scenarios. We assume two different merger models: \emph{MD without delay} (solid lines) and \emph{Dominik} (dashed lines). The error bands around the curves represent uncertainties in merger rates as obtained from GWTC-3 data. The timeline is segmented into observing runs and detector upgrades, highlighted with shaded regions. \emph{Bottom panel}: The percentage of lensed events among the BBH detections. Since the Dominik merger rate has relatively higher support at higher redshift, it results in higher lensing fraction. The non-monotonic behavior of the lensing fraction is due to the selection effects of the detector networks (inability to detect one of the images due to gaps between observing runs).
}
\label{fig:lfraction_and_mrate}
\end{figure*}

If the start and end time of each observing segment $i$ of the detector network is given by $T^s_i$ and $T^e_i$ respectively (see Fig.~\ref{fig:revised_duty_cycle}), Eq.~\eqref{eq:snr_dist_joint} can be simplified as 
\begin{align}
\frac{d^2\tilde{P}}{d\rho_1 d\rho_2}(\zs, \Delta t) &= \frac{1}{T_d} \sum_{i=1}^{K} \int_{T^s_i}^{T^e_i} dt ~ D(t+\Delta t) \nonumber \\    
&\times\int d\vtheta ~ \dfrac{dP}{d\vtheta} \dfrac{d^2P}{d\rho_1 d\rho_2} (\zs, \vtheta, t, \Delta t), 
\label{eq:snr_dist_joint_simple}
\end{align}
where $K$ is the total number of observation segments. 

Similar to Eq.~\eqref{eq:selection_fn_unlens}, the selection function for lensed events can be found by integrating this joint distribution using the SNR threshold as the lower limit. However, here we should consider the fact that the SNRs of the two lensed events are magnified by factors $\mu_+^{1/2}$ and $\mu_-^{1/2}$, which in turn depend on the impact factor $y$ (see the bottom panel of Fig.~\ref{fig:selection_fn_inputs}). Thus, 
\begin{equation}
S^L(\zs, \Delta t, y ) = \int_{\rho_\mathrm{th}/\mu_+^{1/2}(y)}^\infty \!\!\! d\rho_1  \int_{\rho_\mathrm{th}/\mu_-^{1/2}(y)}^\infty \!\!\! d\rho_2 \, \frac{d^2\tilde{P}}{d\rho_1 d\rho_2}(\zs, \Delta t). 
\label{eq:sel_fun_lens_zs_dt_y}
\end{equation}
Now the selection function can be marginalized over $y$ and $\Delta t$. 
\begin{equation}
S^L(\zs)  =\int_0^1 dy \frac{dP}{dy} \int_0^{T_d} d\Delta t \frac{dP}{d\Delta t} (\zs, y) ~ S^L(\zs, \Delta t, y ),
\label{eq:lens_sel_fun_sz}
\end{equation}
where the obvious prior for $y$ is uniform on a unit disk, i.e: $dP/dy \propto y$ while   
\begin{equation}
\frac{dP}{d\Delta t} (\zs, y) =  \int_0^{\zs} d\zl \int_{\sigma_\mathrm{min}}^ {\sigma_\mathrm{max}} d\sigma ~ \frac{d^2P}{d\sigma d\zl}(\zs) ~ \frac{dP}{d\Delta t} (\zs, \zl, \sigma, y). 
\label{eq:time_delay_dist_fn_zs_y}
\end{equation}
Above, ${d^2P}/{d\sigma d\zl}$ is obtained from the differential optical depth ${d^2 \tau}{/d\sigma d\zl}$ (the integrand in Eq.~\eqref{eq:SIS_optd}), while ${dP}/{d\Delta t} (\zs, \zl, \sigma, y)$ is given by the Dirac Delta function at $\Delta t (\zs, \zl, \sigma, y)$.
\begin{equation}
\frac{dP}{d\Delta t} (\zs, \zl, \sigma, y) = \delta[\Delta t - \Delta t(\zs, \zl, \sigma, y)],
\label{eq:dirac_delta}
\end{equation}
where $\Delta t(\zs, \zl, \sigma, y)$ is given in Eq.~\eqref{eq:deltat-SIS}.

\subsection{Lensing detection rates for future observations}

Putting all these together, we estimate the expected (cumulative) number of total and lensed BBH detections ($\Lambda$ and $\Lambda_l$) in upcoming observing scenarios as a function of the observation time (top panel of Fig.~\ref{fig:lfraction_and_mrate}). The expected cumulative detection fraction of lensed events $u_\mathrm{det} \equiv \Lambdal/\Lambda$ is shown in the bottom panel of Fig.~\ref{fig:lfraction_and_mrate}. We present these predictions for \texttt{Planck18} cosmology, and we show in Sec.~\ref{sec:bayes_formalism} how these estimates vary with cosmology.   

The reader might notice the non-monotonic nature of the lensing fraction as a function of the observation time. This might look surprising, given that the horizon distance of the detectors is expected to increase in the future (due to better sensitivity) and the lensing optical depth is larger at higher redshifts (see Fig.~\ref{fig:optd_hmf_sdss}). The non-monotonic nature of the \emph{observed} lensing fraction is due to the selection effects in detecting lensed events. In order to detect a lensed event, both of its images need to be detected. If one of the images happens to arrive at the detector when it is not operational, we will miss that event. This effect is particularly prominent during the gaps between major observing runs. This is why we see a drop in the observed lensing fraction when a new observation run is started after a gap.

\subsection{Time delay distribution of detectable lensed events}
\label{sec:time-delay_dist}
\begin{figure}[t]
	\centering
	\includegraphics[width=\columnwidth]{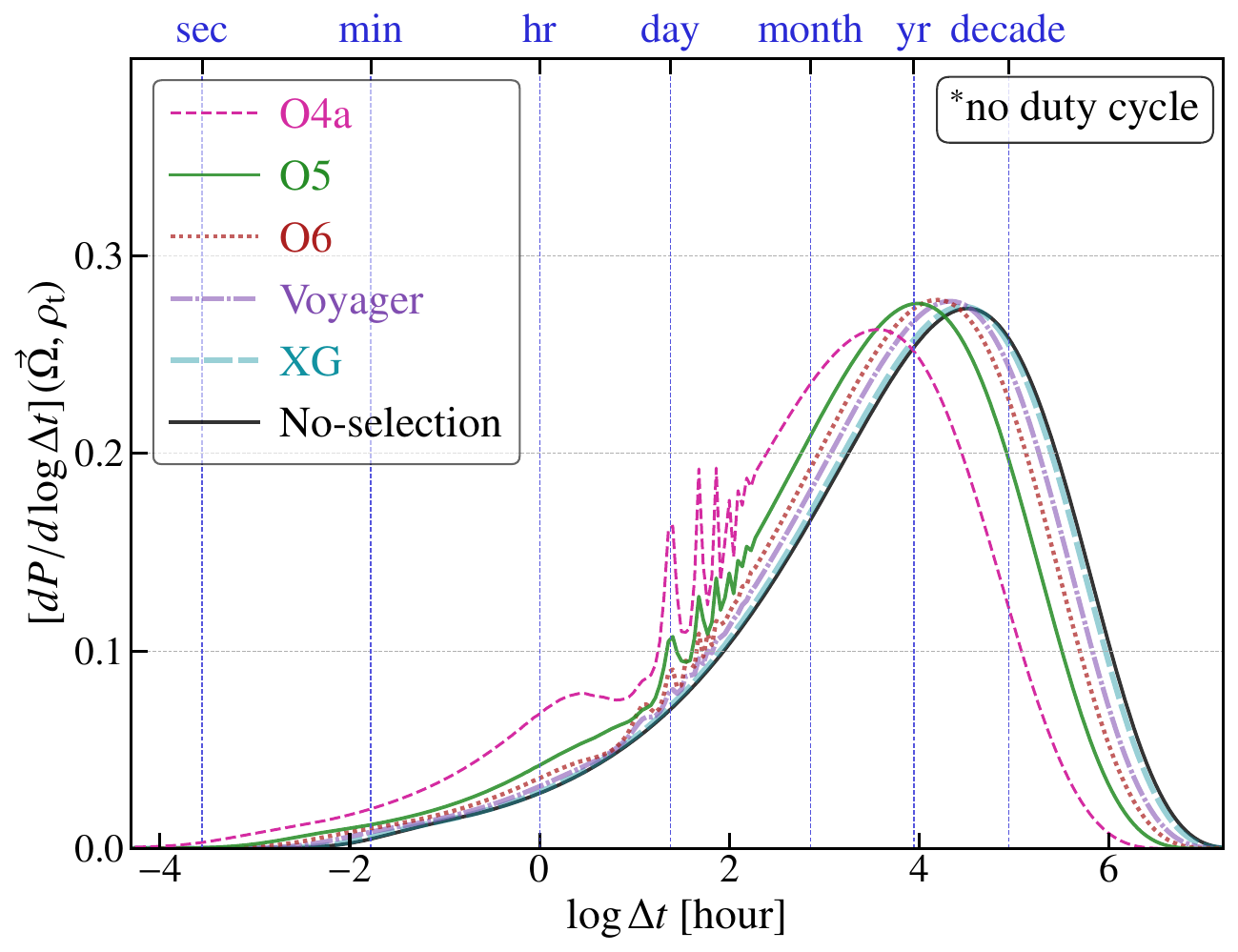}
	\caption{Time-delay distributions of lensed events detectable in various observing scenarios, assuming \texttt{Planck18} cosmology and \emph{Dominik} merger-rate distribution (Fig.~\ref{fig:merger_rate_all_models}, left). As the detector network sensitivity improves, the time-delay distribution shifts towards the right, approaching the distribution in the absence of selection effects.}
	\label{fig:td_dist_all_detect_scenario}
\end{figure}

In Eq.~\eqref{eq:time_delay_dist_fn_zs_y}, we had obtained the intrinsic time-delay distribution after marginalizing over lens parameters. To obtain the time-delay distribution of the detected lensed events, we marginalize over the expected distributions of the source redshift and the impact parameter while incorporating the corresponding selection function $S^L(\zs, \Delta t, y)$ from Eq.~\eqref{eq:sel_fun_lens_zs_dt_y}.
\begin{align}\label{eq:td_bias_master}
\dfrac{dP^{\rm det}}{d\Delta t}& \propto \int_0^{z_\mathrm{max}} \!\!\!\! d\zs \dfrac{dP_{b}}{d\zs} P_{\ell}(\zs) \int_0^1 \!\!\!\! dy \frac{dP}{dy}~ S^L(\zs, \Delta t, y) ~ \frac{dP}{d\Delta t} (\zs, y)
\end{align}
where $dP_{b}/d\zs$ is the distribution of the intrinsic merger rate (the integrand of   Eq.~\eqref{eq:int_BBH_rate}):
\begin{align}\label{eq:lensed_mrate}
	\dfrac{dP_{b}}{d\zs} (\zs) &\propto \dfrac{1}{(1 + \zs)} ~ \dfrac{d^{2}N}{dV_{c}dt_{\rm s}}(\zs)~\dfrac{dV_{c}}{d\zs}(\zs).
\end{align}

An interesting consequence of including selection effects is the emergence of a modulation in the time-delay distribution with a period of one sidereal day. This arises because the lensing selection function $S^L(\zs, \Delta t, y)$ given in Eq.~\eqref{eq:sel_fun_lens_zs_dt_y} exhibits the same periodic variation due to Earth's rotation affecting the antenna patterns. At larger time-delays, this modulation gets averaged out. Furthermore, the amplitude of these oscillations in $S^L(\zs, \Delta t, y)$ decreases for more sensitive detector networks, causing this feature to diminish as we progress from O4a to XG detectors.  

Figure~\ref{fig:td_dist_all_detect_scenario} shows the time-delay distributions of lensed events detectable in various observing scenarios. As detector network sensitivity improves, these distributions shift toward larger values of $\Delta t$, approaching the intrinsic time-delay distribution. This trend can be understood from Eq.~\eqref{eq:deltat-SIS}: holding other parameters fixed, as source redshift increases, the ratio of angular diameter distances $D_{\ell} D_{\ell s}/D_{s}$ increases, which in turn increases $\Delta t$. Consequently, improved sensitivity enables us to probe higher redshifts, leading to the detection of a larger fraction of lensed events from distant sources with correspondingly larger time delays. Note that Eq.~\eqref{eq:td_bias_master} reduces to Eq.~(17) of Ref.~\cite{Jana_2024} in the limit $\rho_{\rm th} = 0$.

\section{Bayesian inference of cosmological parameters}
\label{sec:bayes_formalism}

\subsection{The imprint of cosmological parameters on the number of strongly lensed pairs and the time delay distribution}

\begin{figure}[tbh]
\centering
\includegraphics[width=\columnwidth]{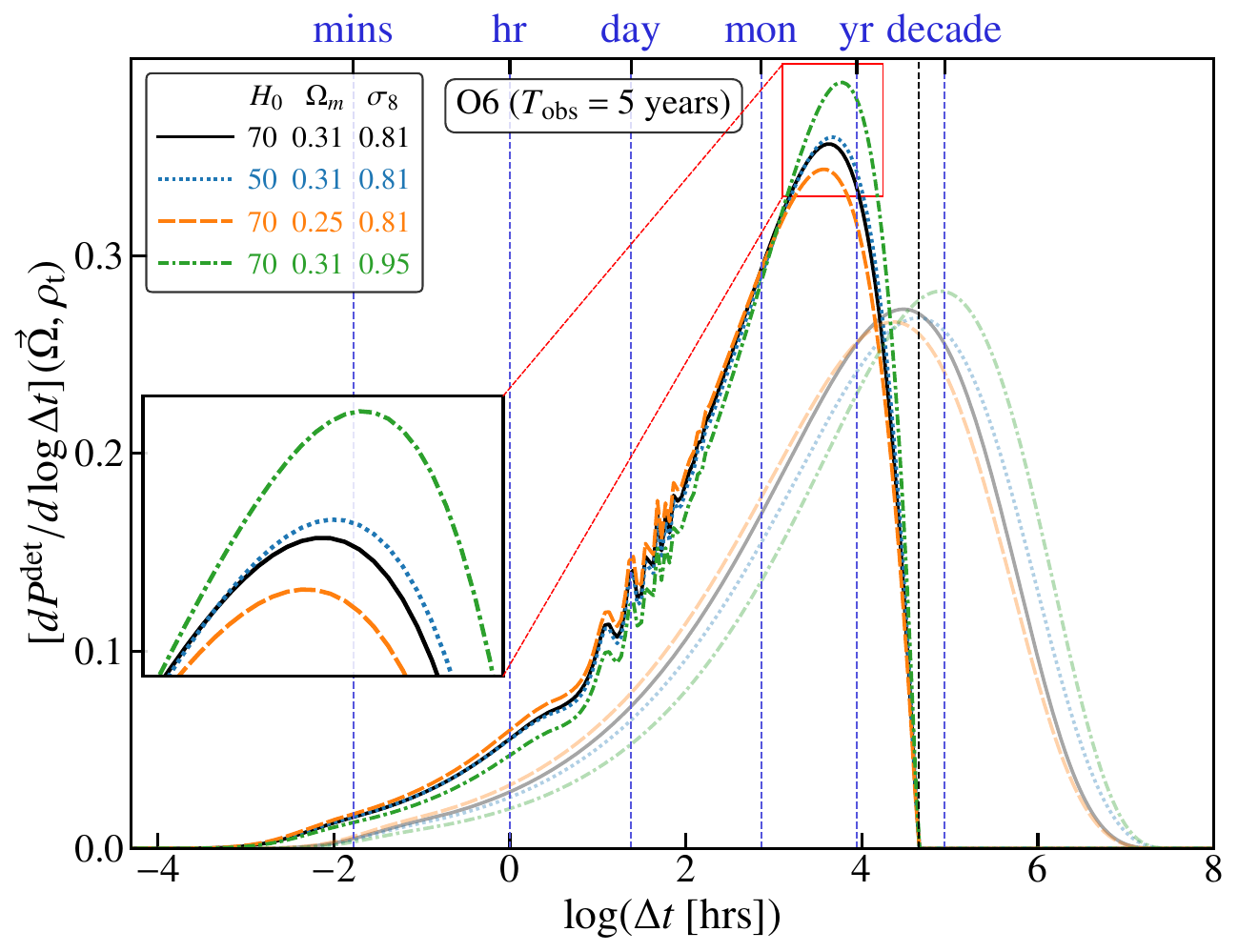}
\caption{Expected variation in the time-delay distribution of lensed events for different values of cosmological parameters $\vOmega \equiv \lbrace H_{0}, \Omega_{m}, \sigma_{8}\rbrace$. We consider the lensed pairs detectable in O6, assuming Dominik merger rate distribution. Our fiducial cosmology  ($H_{0} = 70$ km s$^{-1}$ Mpc$^{-1}$, $\Omega_{m} = 0.31$, $\sigma_{8} = 0.81$) is represented by the black solid curve. As we decrease $H_{0}$, the distribution peaks at larger $\Delta t$. In contrast, the distribution peaks at higher $\Delta t$ as we increase $\Omega_{m}$ and $\sigma_{8}$. 
Light curves are the corresponding distributions without including the effect of the selection function.}
\label{fig:td_temp_var_with_cosmo}
\end{figure}

\begin{figure*}[tbh]
\centering
\includegraphics[width=\textwidth]{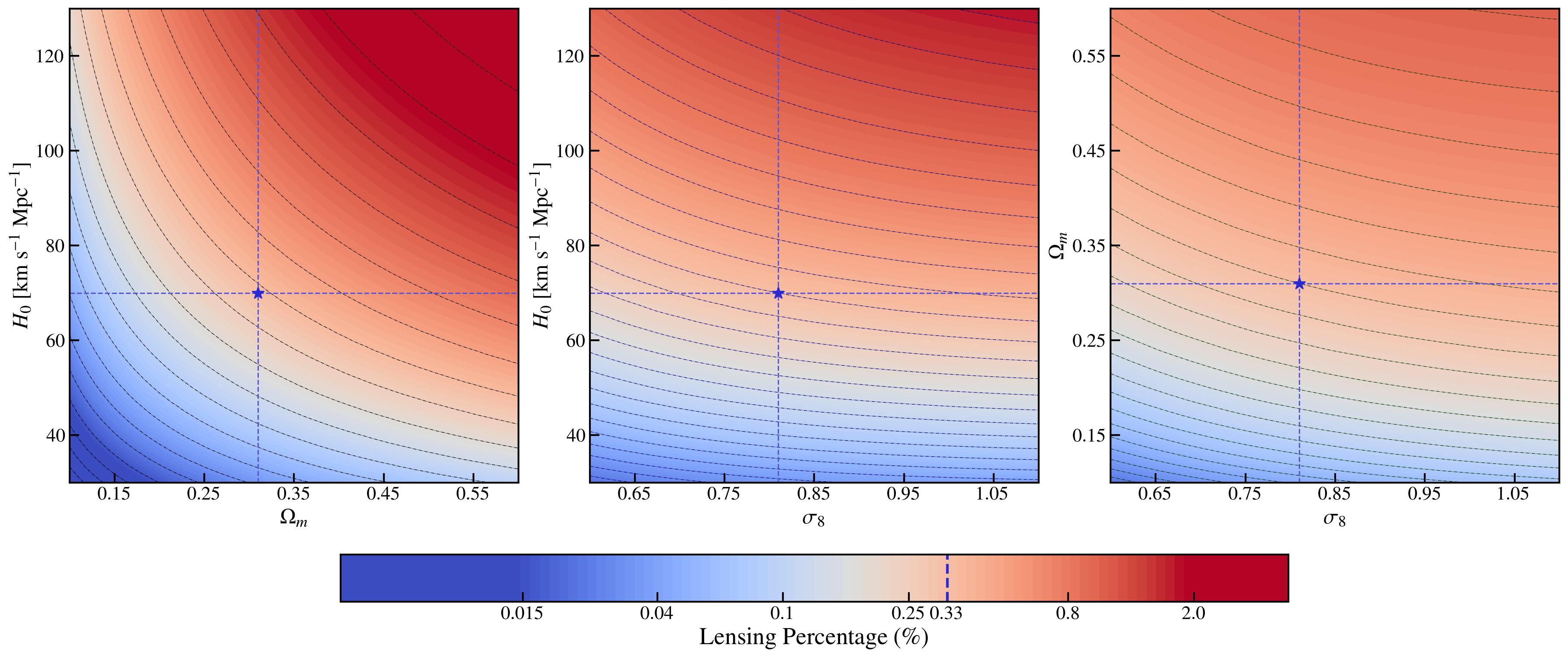}
\caption{Detectable lensing fraction as a function of cosmological parameters for a single observing run. We vary two parameters simultaneously while holding the third fixed: (a) $H_{0}$ and $\Omega_{m}$ with $\sigma_{8} = 0.81$, (b) $H_{0}$ and $\sigma_{8}$ with $\Omega_{m} = 0.31$, and (c) $\Omega_{m}$ and $\sigma_{8}$ with $H_{0} = 70~\text{km}~\text{s}^{-1}~\text{Mpc}^{-1}$. These results assume O6 network sensitivity and a BBH merger rate following the \textit{Dominik} distribution. The crosshairs in all panels mark the fiducial cosmology used for signal injection in our cosmological inference analyses.}
\label{fig:lfrac_var_with_cosmo}
\end{figure*}

In Sec.~\ref{sec:strongly_lensed_expectation}, we computed the expected number of total and lensed BBH mergers assuming a cosmology. These numbers could vary depending on the cosmological model or parameters. Considering the flat $\Lambda$CDM model, its parameters $\vOmega$ can affect our results in two different ways. 

First, they determine the relation between various distance measures (e.g., angular diameter distance, luminosity distance, etc.) and the cosmological redshift, i.e., $\Dl(\zl , \vOmega)$, $\Ds(\zs , \vOmega)$, $\Dls(\zl, \zs , \vOmega)$, etc. As a result: 
\begin{itemize}
\item The redshift distribution of GW sources in Eqs.~\eqref{eq:int_BBH_rate}, \eqref{eq:R_det}, \eqref{eq:exp_ind_det}, \eqref{eq:exp_nl_master_eq} as well as the comoving volume element become functions of cosmological parameters:  
$$\frac{d^{2}N}{dV_{c}dt_{s}}(\zs) \rightarrow \frac{d^{2}N}{dV_{c}dt_{s}}(\zs , \vOmega) ~~ \mathrm{and} ~~ \frac{dV_c}{dz}(z) \rightarrow \frac{dV_c}{dz}(z, \vOmega).$$

\item The detector network selection functions in Eqs.~\eqref{eq:selection_fn_unlens} and \eqref{eq:lens_sel_fun_sz} are functions of the luminosity distance to the sources. In terms of the source {redshift}, then, these become dependent on the cosmological parameters:  
$$S(\zs) \rightarrow S(\zs , \vOmega) ~~\mathrm{and}~~ 
S^{L}(\zs) \rightarrow S^{L}(\zs , \vOmega).$$

\item  Einstein radius in Eq.~\eqref{eq:einstein_radius} becomes a function of cosmological parameters:  
$$r_E(\zl, \zs) \rightarrow r_E(\zl, \zs , \vOmega).$$ 
\item The lensing time delay in Eq.~\eqref{eq:deltat-SIS} becomes a function of cosmological parameters:
$$ \Delta t (\zl, \sigma, \zs, y) \rightarrow \Delta t (\zl, \sigma, \zs, y , \vOmega).$$
\end{itemize}
Second, cosmological parameters affect the structure formation, thus affecting the number density of lenses in Eq.~\eqref{eq:num_dens_lens}: 
$$\frac{d^2n}{d\sigma dV_{c}}(\zl) \rightarrow \frac{d^2n}{d\sigma dV_{c}}(\zl , \vOmega).$$
 
As a consequence of these two broad effects, the differential optical depth and hence the optical depth, in Eqs.~\eqref{eq:diff_opt_depth} and \eqref{eq:opt_depth} respectively, become functions of the cosmological parameters, as do the time-delay distribution in Eq.~\eqref{eq:time_delay_dist_fn_zs_y} and the lensing selection function in Eq.~\eqref{eq:lens_sel_fun_sz}:
\begin{eqnarray}
 \dfrac{d\tau}{d\zl} (\zl, \zs)  &\rightarrow& \dfrac{d\tau}{d\zl} (\zl, \zs, \vOmega) ~~~ \mathrm{and} ~~~  \tau(\zs) \rightarrow \tau(\zs, \vOmega). \nonumber \\
\frac{dP}{d\Delta t} (\zs, y) &\rightarrow& \frac{dP}{d\Delta t} (\zs, y, \vOmega) ~~~ \mathrm{and} ~~~ S^L(\zs)  \rightarrow S^L(\zs, \vOmega).   \nonumber 
\end{eqnarray}

The final result of all these is that the expected number of total detectable (\textit{independent}) events as well as the number of detectable lensed pairs, along with their time-delay distribution, contain imprints of cosmological parameters:
\begin{eqnarray} 
	\Lambda   &\rightarrow& \Lambda (\vOmega)  \nonumber \\ 
	\Lambdal  &\rightarrow& \Lambdal (\vOmega) \nonumber \\
	\dfrac{dP^{\rm det}}{d\Delta t} &\rightarrow& \dfrac{dP^{\rm det}}{d\Delta t} (\vOmega). \nonumber
\end{eqnarray}

In Fig.~\ref{fig:td_temp_var_with_cosmo}, we plot the time-delay distribution of the strongly lensed pairs detectable in $T_{\rm obs} = 5$ year observing run in {O6}, demonstrating its sensitivity to varying cosmological parameters. The peak exhibits parameter-dependent shifts: increasing $H_{0}$ moves the peak towards smaller values, while rising both $\Omega_{m}$ and $\sigma_{8}$ drive it towards larger time-delay values. Figure~\ref{fig:lfrac_var_with_cosmo} shows how the detectable lensing fraction varies with cosmological parameters. Below, we employ these observables to show how to constrain the cosmological parameters. 

\subsection{Likelihoods from lensing fraction and time-delays}

\begin{figure*}[tbh]
\centering 
\includegraphics[width=0.97\textwidth]{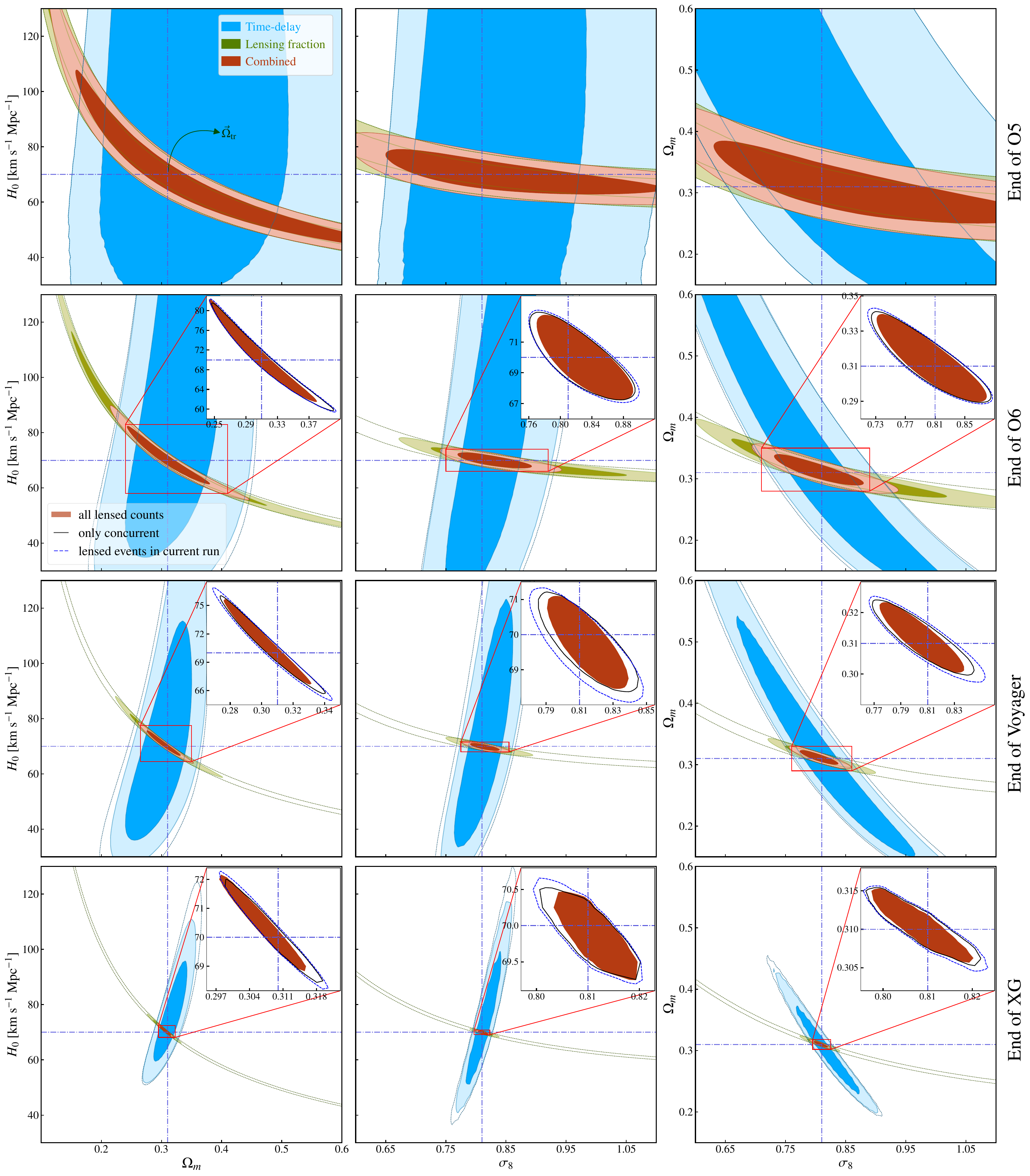}
\caption{Expected posterior distributions of \(H_0, \Omega_m\), and \(\sigma_8\) at the end of different observing runs (O5, O6, Voyager, and XG). Each panel shows \(68\%\) (darker shade) and \(95\%\) (lighter shade) credible regions from different lensing observables: lensing fraction (olive), time-delay distribution (blue), and combined constraints (brown). Results are shown for one realization of the observing scenario, assuming \textit{Dominik} merger rate distribution, normalized to the median rate of GWTC-3 at low redshifts. Dashed contours in the main figures represent \(95\%\) credible regions derived from those lensed pairs for which both the images can be detected in that particular observing run. For O6, Voyager, and XG (shown in the insets), we display the \(68\%\) credible interval of the combined posterior for three scenarios: (a) shaded: all lensed events detected up to that observing run, (b) solid: only concurrent pairs up to that observing run (see text), (c) dotted: lensed events from only the current observing run.}
\label{fig:all_combined_2D_posteriors}
\end{figure*}

\begin{figure*}[tbh]
\centering
\includegraphics[width=\textwidth]{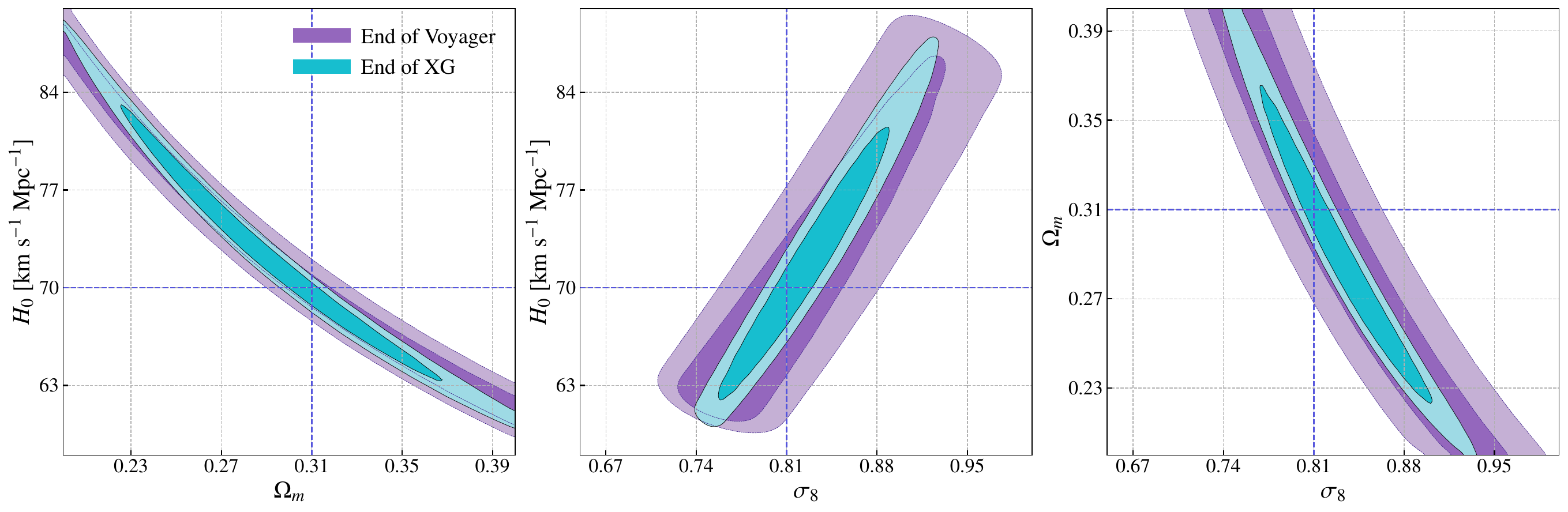}
\caption{Representative posterior distribution when all three cosmological parameters are varied simultaneously, showing $68\%$ and $95\%$ credible regions for the combined likelihood. Only lensed events detectable in Voyager and XG are included, assuming \textit{Dominik} merger rate distribution normalized to the median merger rate of GWTC-3 at low redshifts.}
\label{fig:3D_posteriors}
\end{figure*}

\begin{figure*}[tbh]
\centering
\includegraphics[width=0.75\textwidth]{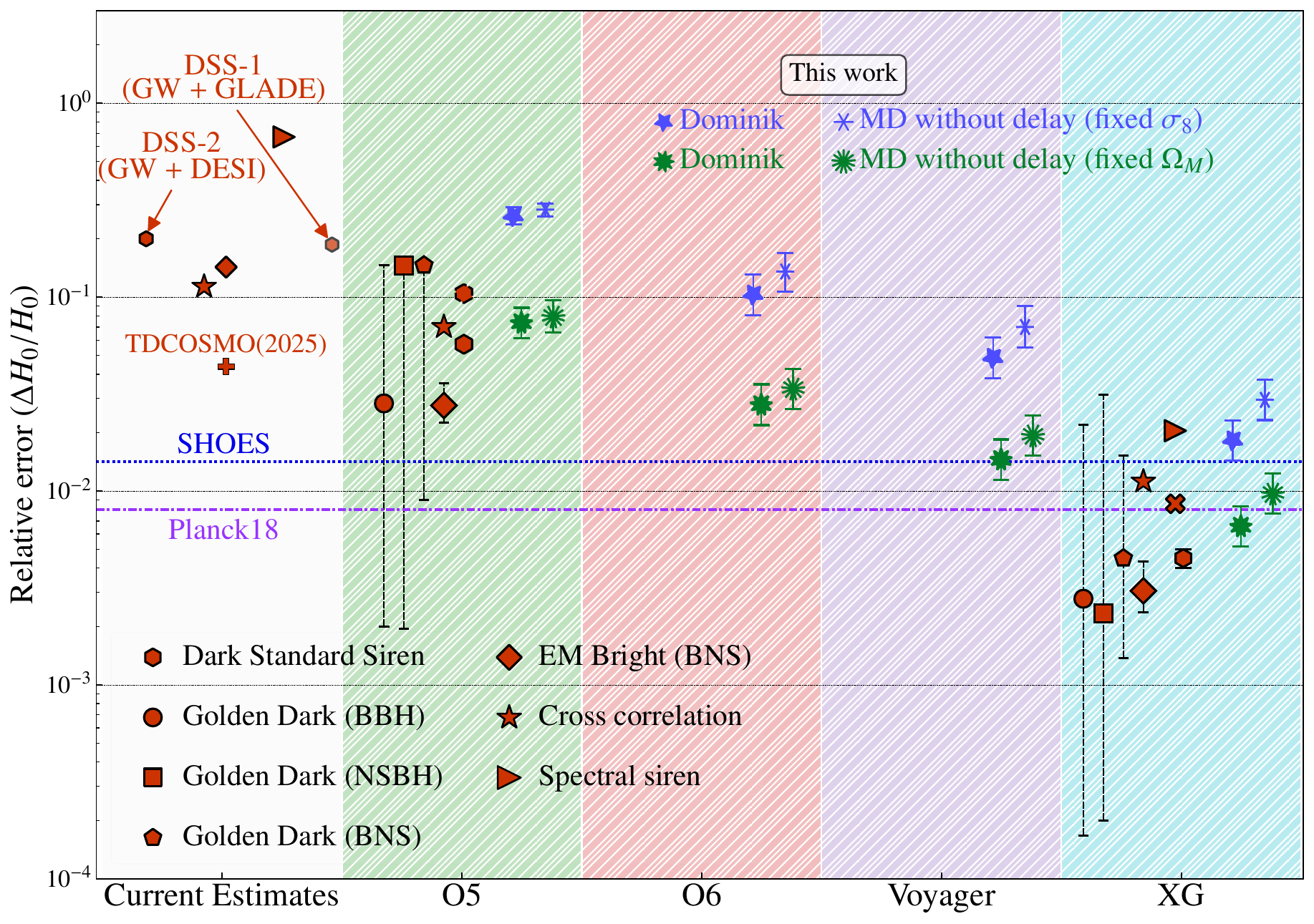}
\caption{\small Current and anticipated measurement errors in Hubble constant ($\Delta H_{0}/H_{0}$) using various methods across different observation periods. The current uncertainties using supernovae (SHOES~\cite{Riess_2022}) and CMB (Planck18~\cite{planck18_A&A} ) are shown by horizontal lines. {TDCOSMO-2025} represents constraints from strong gravitational lensing (optical) studies~\cite{birrer2025tdcosmo}. The diamond  shows the precision achieved from the multi-messenger observation of the binary neutron star merger GW170817~\cite{abbott2017_bright_siren}. Two dark standard siren approaches are indicated: {DSS-1} combines 46 dark sirens from O1-O2-O3 with the GLADE galaxy catalog~\cite{dalya2018glade, dalya2022glade+}, while {DSS-2} utilizes 8 dark sirens from O1-O2-O3 with the DESI galaxy survey~\cite{palmese2023standard}. The spectral siren technique was applied to 42 BBH events from GWTC-3~\cite{abbott_2023a}. Finally, the cross-correlation method employs 8 well-localized GWTC-3 events combined with photometric surveys from 2MPZ and WISE-SuperCOSMOS~\cite{mukherjee2024cross}.  For predictions in O5 and XG using golden dark sirens and bright sirens, we adopt results from~\cite{chen2024cosmography}. For one year of observation in each scenario, it provides constraints across different binary systems (BBH, NSBH, and BNS). The hexagons in second column show the predictions using dark sirens in O5, assuming two different scenarios on the sky localization (see Table I~\cite{alfradique2025systematic}). In the XG era, the precision for dark sirens is obtained from~\cite{muttoni2023dark}, assuming different completeness for the galaxy catalog, for one year of observation. In the XG era, we show precision estimates from spectral siren measurements alone (arrow marker in the rightmost column) as well as from joint inference combining spectral and (golden) dark sirens (the cross marker). We also include projections from the cross-correlation approach~\cite{afroz2024prospect} (the star markers in second and last column). Finally, we present forecasts for time-delay cosmography in upcoming observing runs under two parameter configurations. When varying $H_{0}$ and $\Omega_{m}$ jointly (with $\sigma_8$ fixed), the $\Omega_{m}$-marginalized $H_{0}$ uncertainty is shown in blue markers. When varying $H_{0}$ and $\sigma_{8}$ jointly (with $\Omega_{m}$ fixed), the $\sigma_{8}$-marginalized $H_{0}$ uncertainty is shown in green markers. For both cases, different star-like symbols represent the \textit{Dominik} merger rate and sunburst symbols represent \textit{MD without delay}.}	
\label{fig:errors_competative}
\end{figure*}

We assume that $N$ BBH mergers are detected, out of which $N_\ell$ events are confidently identified to be strongly lensed. We further assume that the time delay between $N_\ell$ image pairs has been measured with negligible errors, yielding $N_\ell$ samples of lensing time delays, i.e., $\{\Delta t_i\}$. We aim to constrain cosmological parameters using $u_\mathrm{det} \equiv N_\ell/N$ and $\{\Delta t_i\}$. The posterior distribution for cosmological parameters $\vOmega$ can be written in the following way using Bayes' theorem 
\begin{align}\label{eq:bayes-Master}
	p(\vOmega|~u_\mathrm{det}, \lbrace \Delta t_{i}\rbrace) = \dfrac{p(\vOmega) ~ p(u_\mathrm{det}, \lbrace \Delta t_{i}\rbrace|~\vOmega)}{Z}
\end{align}
where $p(\vOmega)$ represents our prior information of the cosmological parameters of interest, which we assume to be {flat}, while $Z$ is the normalisation constant, which can be interpreted as the evidence of the model (which includes other aspects such as the HMF model). Since the lensing fraction and the observed time-delays are mutually independent, the likelihood can be split as follows:
\begin{equation}\label{eq:lhd-Master}
	p(u_\mathrm{det}, \lbrace \Delta t_{i}\rbrace|~\vOmega) = p(u_\mathrm{det}|~\vOmega) ~ p(\lbrace \Delta t_{i}\rbrace|~\vOmega).
\end{equation}
Note that this formulation is the same as that presented in Jana et al~\cite{Jana_2023,Jana_2024}, except that here the likelihood is defined on the lensing fraction $u_\mathrm{det}$ instead of the detected number of lensed signals $N_\ell$.

\subsubsection{Likelihood of the measured lensing fraction} 
Since BBH merger events occur independently, their detection counts $N$ follow a Poisson distribution with mean $\Lambda$. Since a fraction of them are lensed, their counts $N_\ell$ follow a Poisson distribution with mean $\Lambda_\ell$. That is, 
\begin{equation}
p(N | \Lambda) = \frac{\Lambda^{N}\;e^{-\Lambda}}{N!} ~~\mathrm{and} ~~~ p(N_\ell | \Lambda_\ell) = \frac{\Lambda_\ell^{N_\ell}\;e^{-\Lambda_\ell}}{N_\ell!} 
\label{eq:N_Poisson_dist}
\end{equation}
We calculate $\Lambda (\vOmega)$ and $\Lambdal(\vOmega)$ using Eqs.~\eqref{eq:R_det} and \eqref{eq:exp_nl_master_eq}  respectively. The likelihood for the lensing fraction $u_\mathrm{det}$ (the probability of observing a fraction $u_\mathrm{det}$ of the events lensed given the cosmological parameters $\vOmega$) is therefore written as a convolution of those two Poisson distributions.
\begin{align}\label{eq:lfrac_lhd_concurrent}
p(u_\mathrm{det} | \vOmega) & =  \int_{0}^{\infty} \!\! dN \int_{0}^{N} \!\! dN_\ell \, \delta \left(u_\mathrm{det} - \dfrac{N_\ell}{N}\right) \nonumber \\
&\times p\left(N |\Lambda (\vOmega)\right) p\left(N_\ell | \Lambdal(\vOmega) \right), 
\end{align} 
where $p\left(N |\Lambda (\vOmega)\right)$ and $p\left(N_\ell | \Lambdal(\vOmega)\right)$ are given by Eq.~\eqref{eq:N_Poisson_dist}, while $\delta$ denotes the Dirac Delta.

\subsubsection{Likelihood of the measured time delays}

Since the individual lensed events are statistically independent, the likelihood of observing a set of lensing time delays $\{\Delta t_i\}$ can be written as the product of the likelihoods of individual time delays. The time delays are measured with millisecond precision (very small as compared to expected lensing time delays), hence the likelihood of observing a time delay $\Delta t_i$ is given by ${dP^{\rm det}}/{d\Delta t}$ (see Eq.~\eqref{eq:td_bias_master} and Fig.~\ref{fig:td_temp_var_with_cosmo}) evaluated at $\Delta t_i$. Therefore, given the cosmological parameters $\vOmega$, the likelihood of observing a set of time delays $\{\Delta t_i\}$ is given by

\begin{equation}\label{eq:td_lhd_eq1}
p\left(\lbrace \Delta t_i\rbrace|~ \vOmega \right) = 
\prod_{i=1}^{N_{\ell}} \frac{dP^{\rm det}}{d\Delta t}\left(\Delta t_i~|~\vOmega \right). 
\end{equation}
           
\section{Expected constraints from strong lensing cosmography}\label{sec:results}

To derive expected constraints on cosmological parameters using lensing observables, we follow the approach of~\cite{Jana_2023,Jana_2024}. The only differences are that here we include the detector network selection effects and use a likelihood on the lensing \emph{fraction} instead of the \emph{number} of lensed events. We create observations using our fiducial cosmology $\vOmega^\mathrm{tr}$. That is, for each observing run (see Table~\ref{tab:networks}), we compute the expected number of detectable BBH mergers $\Lambda(\vOmega^\mathrm{tr})$ and lensed events $\Lambda_\ell(\vOmega^\mathrm{tr})$ using Eqs.~\eqref{eq:R_det} and \eqref{eq:exp_nl_master_eq}, respectively. We perform Poisson sampling to get one value of $N$ and $N_\ell$, that we call $N^{\rm tr}$ and $N^{\rm tr}_\ell$. We then draw $N_\ell^{\rm tr}$ time-delay samples from the distribution ${dP^{\rm det}}/{d\Delta t}$ assuming our fiducial cosmology $\vOmega^{\rm tr}$ using Eq.~\eqref{eq:td_bias_master}. We next calculate both the model time-delay distributions and the expected lensing fraction for different choices of $\vOmega$. Finally, we calculate the corresponding likelihoods following Eq.~(\ref{eq:lhd-Master}). We use a flat prior on $\vOmega$ in Eq.~(\ref{eq:bayes-Master}) to obtain the posterior on different cosmological parameters.

Figure~\ref{fig:all_combined_2D_posteriors} shows constraints on $H_{0}, \Omega_{m}$, and $\sigma_8$ obtained at the end of four observing runs: O5, O6, Voyager, and XG. Each panel shows the posteriors on two cosmological parameters while fixing the third to its fiducial value (similar to Fig.~\ref{fig:lfrac_var_with_cosmo}). For each run, we display three likelihood contours (same as posteriors as we use flat priors): one derived from the lensing fraction, one from time-delay samples, and their combination. The HMF used to model the lens population is adopted from~\cite{Behroozi_2013}, while the unlensed merger rate distribution follows the Dominik~\cite{Dominik_2013} model, calibrated to the median merger rate distribution inferred from GWTC-3 (Fig.~\ref{fig:merger_rate_all_models}). 

Beyond O5, all posteriors incorporate lensed events detected across multiple observing runs. The filled contours represent constraints using all lensed pairs accumulated up to each run. For comparison, dashed contours show constraints obtained using only concurrent lensed pairs --- those where both images are detected within the same observing run. The inset panels (excluding top left) display posteriors obtained when deliberately excluding such cross-run pairs (solid contours in insets), showing the importance of including such events. All posteriors get tighter as detector network  sensitivity improves across observing runs --- increased lensing statistics directly translate to improved cosmological constraints. Figure~\ref{fig:3D_posteriors} shows the posteriors on the same cosmological parameters, after \textit{marginalizing} over the third parameter. Here, due to the computational cost, we show the results only for the Voyager and XG detectors. 

In Fig.~\ref{fig:errors_competative}, we compare the expected constraints on $H_{0}$ using lensing cosmography with the same expected from GW standard sirens. Different kinds of GW standard sirens are proposed in the literature. The most straightforward approach relies on identifying an electromagnetic (EM) counterpart and its host galaxy (``bright sirens'')~\cite{Holz_2005, Dalal_2006, Nissanke_2009, 2019ApJ_Fishbach}. When no EM counterpart is available, we can perform statistical host identification, weighting each galaxy within the sky localization posterior appropriately (``dark sirens'')~\cite{Schutz_1986Nature, Del_Pozzo_2012, Nair_2014, Soares_Santos_2019, gray_2020, Borhanian_2020}. A third method exploits features in the mass distribution of sources, inferring redshift from events at similar luminosity distances by comparing detector-frame and source-frame masses (``spectral sirens'')~\cite{ezquiaga2022spectral, Mali_2024}. For reference, the existing precision in the $H_0$ measurements using Type 1a supernovae and CMB is also shown. Broadly, we see that the expected constraints from lensing cosmography are comparable to that from GW standard sirens.

\section{Summary and Future Work}
\label{sec:outlook_and_conclusion}
In this paper, we investigate the possibility of performing GW strong lensing cosmography~\cite{Jana_2023} using data from upcoming observing runs of current-generation detectors as well as XG detectors. This requires a careful treatment of the selection effects of GW detectors, which were neglected in earlier works~\cite{Jana_2023,Jana_2024}. Note that the selection effects of lensed GW signals are distinct from that of unlensed signals --- detection of a lensed GW pair requires both images to be detected confidently. This warrants a careful consideration of the distribution of the lensing magnification, time delay as well as the observation time of GW detectors, including gaps in the observations. We consider BBH mergers as our GW sources. Their merger rate distribution is modelled using two different astrophysical models, after calibrating them with low-redshift GW observations by LIGO and Virgo. Lenses are modelled as SISs, with parameters inferred from a HMF model. 

Through Bayesian analysis of simulated populations, we demonstrate that measurements of strong lensing observables (fraction of lensed events as well as the lensing time delays) can constrain key cosmological parameters ($H_{0}$, $\Omega_{m}$, $\sigma_{8}$) of the flat $\Lambda$CDM model, with precision improving systematically as more strongly lensed events are accumulated. The projected constraints from lensing cosmography are comparable to those expected from GW standard sirens. In the era of XG observations, the selection effects of GW detectors are not very significant, and our revised forecasts are similar to the earlier forecasts~\cite{Jana_2023,Jana_2024}.

This paper presents an important step toward developing the lensing cosmography concept introduced by Jana et al.~\cite{Jana_2023} into a practical observational program. Previous work addressed systematic errors due to the imperfect modeling of the HMF, imperfect reconstruction of the source properties, and contamination in the catalog of lensed events~\cite{Jana_2024}. The present analysis still employs a simplified SIS lens model and a simple prescription to map the properties of dark matter halo to that of the gravitational lens. In future work, we plan to incorporate more realistic lens models and develop improved prescriptions to connect HMFs with lens properties. 

More realistic lens models, such as the singular isothermal ellipse, will introduce new features, such as the formation of quadruple images~(see, e.g.,~\cite{more2022improved}). This will allow us to form many more pairs from a single lensed event, potentially increasing the information content, leading to better constraints on cosmological parameters. Our formulation for incorporating the selection function in the lensing cosmography analysis can be generalized to such situations. Note that more realistic lens models are likely to change our current estimates of the lensing fraction and time-delay distribution.  However, this does not affect our key point --- the observed lensing fraction and the time delay distribution will contain imprints of cosmological parameters. By appropriately modeling the source and lens population, we will be able to extract cosmological information from the lensing observables. 

Although this method shows strong potential as a complementary cosmological probe --- particularly by accessing the intermediate-redshift regime --- there remains substantial scope for refinement. Here, we have used only the time-delay information from lensed events. Future extensions could include additional observables such as the distribution of the magnification ratio of lensed images~\cite{more2022improved,oguri2018effect}. Finally, if it becomes possible to statistically associate lens galaxies with a subset of strongly lensed events (analogous to the GW dark siren approach), this could provide valuable supplementary information and  enhance cosmological parameter precision. We are currently exploring these directions in ongoing work.

\subsection*{Acknowledgments}
We are grateful to Anupreeta More for useful comments on the manuscript. We also thank the members of the ICTS Astrophysical Relativity group and members of the LVK Lensing group for useful discussions. Our research was supported by the Department of Atomic Energy, Government of India, under Project No. RTI4001. Numerical simulations were performed on the Alice computing cluster at International Centre for Theoretical Sciences, Tata Institute of Fundamental Research.

\appendix

\begin{figure*}[tbh]
\centering
\includegraphics[width=0.85\textwidth]{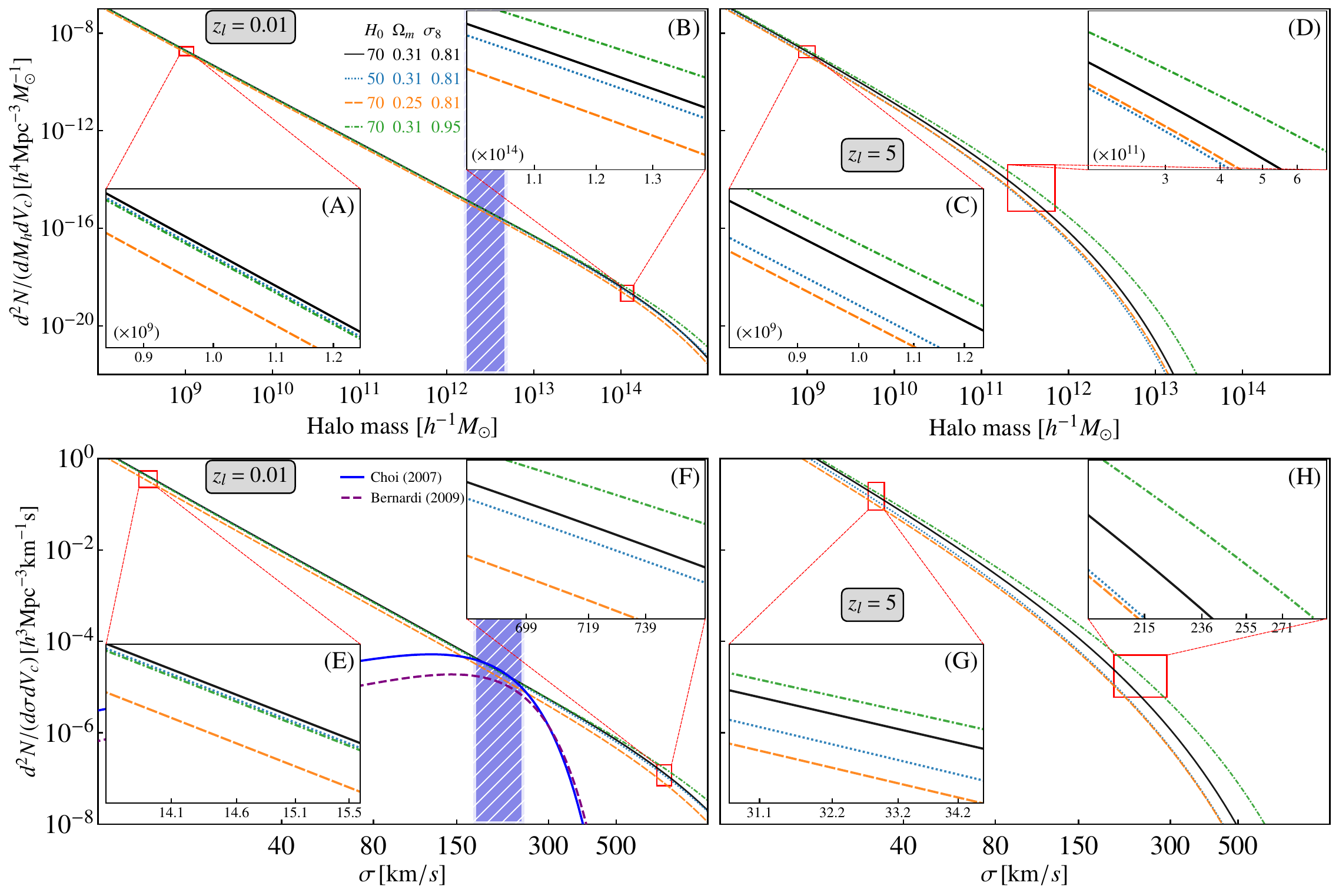}
\caption{\textit{Top}: HMF at lens redshifts $\zl = 0.01$ (left) and $\zl = 5$ (right) for different cosmological parameters, assuming the Behroozi~\cite{Behroozi_2013} model. In the insets, we zoom into two different halo mass intervals (lower and higher mass ranges) to better visualize their cosmological dependencies. \textit{Bottom}: Corresponding velocity dispersion functions (following the prescription of Eq.~\eqref{eq:sigma_prescription}) at the same lens redshifts. The left panel also shows the velocity dispersion function for the lens population distributed according to the SDSS catalog for early-type galaxies (blue and purple curves) in the local universe. The shaded regions in the left column indicate the halo mass interval (top) and velocity dispersion interval (bottom), respectively, where the HMF predictions agree with SDSS observations in the local universe.}
\label{fig:hmf_var_with_cosmo}
\end{figure*}

\begin{figure}[t]
	\centering
	\includegraphics[height=2.3in]{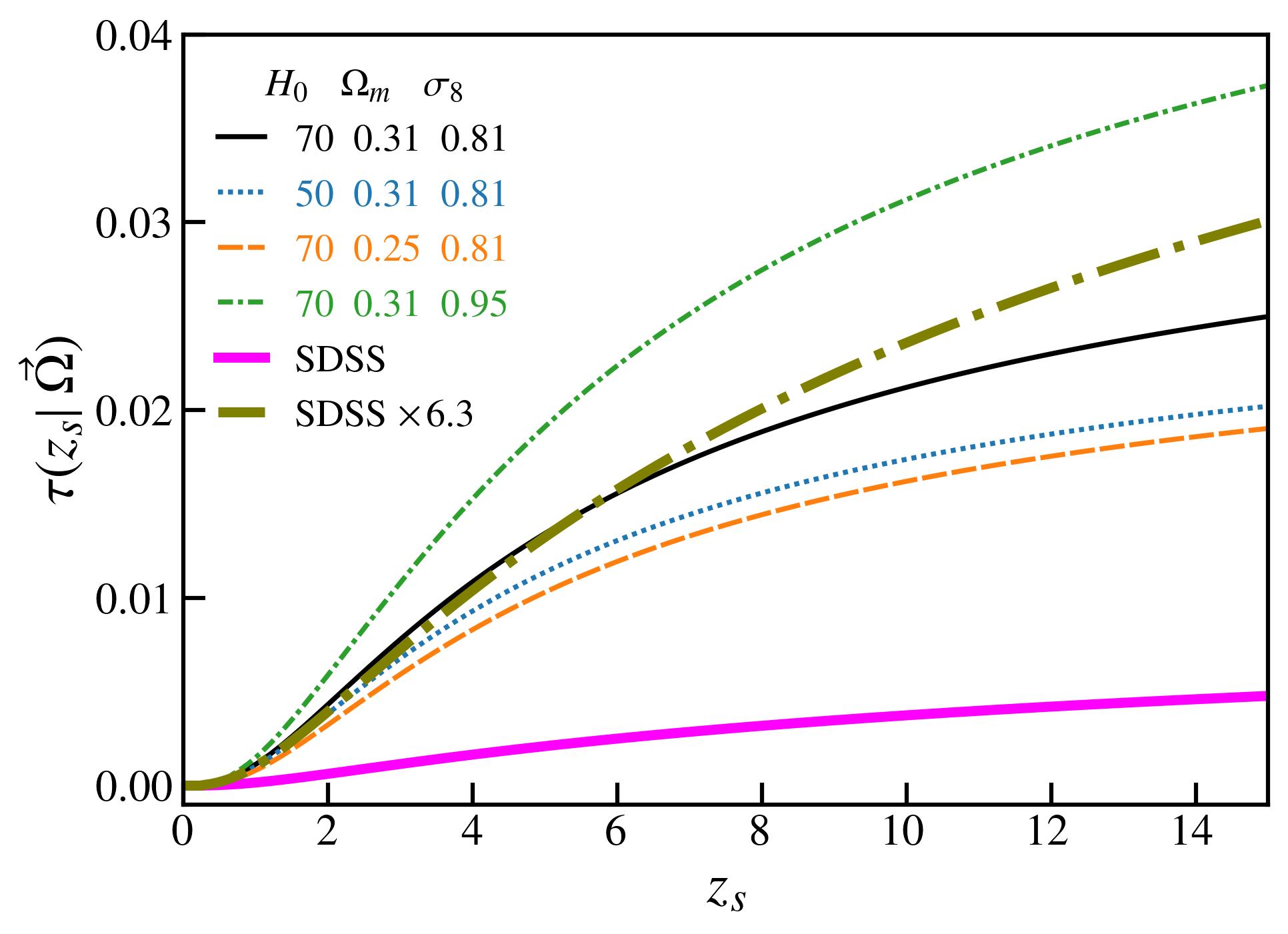}
	\caption{Strong lensing optical depth as a function of source redshift $\zs$. Thick lines show the optical depth for lens populations distributed according to the SDSS catalog of galaxies (magenta and olive), while thinner lines represent the same for lens populations distributed according to the HMF for different cosmological parameters. We use the same color scheme as Fig.~\ref{fig:hmf_var_with_cosmo}. The optical depth exhibits strong dependence on both the cosmological parameters and the lens population model.} 
	\label{fig:optd_hmf_sdss}
\end{figure}

\section{Effect of cosmological parameters on lensing observables}
\label{app:cosmo_effect_on_lensing}

Here we detail the effect of cosmological parameters on the lens population. The top panel of Fig.~\ref{fig:hmf_var_with_cosmo} shows the HMF at two redshifts ($\zl = 0.01, 5$) corresponding to different cosmological parameters. Fixing $H_0$ and $\sigma_8$ to their fiducial values, we observe that increasing $\Omega_m$ results in a higher halo number density across all mass scales. This occurs simply because a higher matter fraction provides more material for structure formation. When we increase $H_{0}$ (fixing other cosmological parameters), the critical density 
($\rho_{c}\propto H_{0}^{2}$) of the universe increases, lowering the mass (or equivalently length) scale for gravitational instability. Therefore, we expect more less-massive halos to form as we increase $H_{0}$.

The impact of $\sigma_8$ on the HMF depends on the redshift. At low redshift ($\zl = 0.01$), the number density decreases for low-mass haloes but increases for high-mass haloes. Although the overall amplitude of fluctuations increases with $\sigma_{8}$, the precise shape of the mass function depends on how the power is distributed across different scales. In this case, when the universe is relatively older, many smaller halos have already merged to form larger structures, depleting the low-mass end. In contrast, at high redshift ($\zl = 5$), when structure formation is still in its early stages, a higher $\sigma_8$ enhances the number density across all halo masses, as mergers have not yet significantly affected the population.

The bottom panel of Fig.~\ref{fig:hmf_var_with_cosmo} shows the corresponding distributions of the velocity dispersion $\sigma$ at same redshifts, following the conversion given in Eq.~(\ref{eq:sigma_prescription}). Relative to the fiducial cosmology (black solid line), all other curves maintain the same trend as the HMFs, which is expected since massive haloes corresponds to higher $\sigma$ at a particular epoch.

For comparison, we also show the $\sigma$ distribution modelled from the SDSS catalog for early-type galaxies in the local universe~\cite{Choi_2007, bernardi2010}. Haris et al.~\cite{haris2018} used the modified Schechter function with parameters fitted to the SDSS early-galaxy catalog~\cite{Choi_2007} to model the expected number density of lenses:
\begin{align}
	\dfrac{d^2n}{d\sigma dV_{c}} = n_{0} \left(\dfrac{\sigma}{\sigma^{*}}\right)^{\alpha} \exp\left[-\left(\dfrac{\sigma}{\sigma^{*}}\right)^{\beta}\right] \dfrac{\beta}{\Gamma(\alpha/\beta)} \dfrac{1}{\sigma},
\end{align} 
where $n_{0} = 8\times 10^{-3} h^{3}/$Mpc$^{3}$, $\alpha = 2.32$, $\beta = 2.67$ and $\sigma^{*} = 161$ km$/$s. 
However, this approach does not account for redshift evolution of the number density following hierarchical structure formation, as the parameters are fitted only to low-redshift observations. Further, it considers only galaxies as gravitational lenses, neglecting the contribution of galaxy clusters as well as low-mass dark-matter halos that do not host galaxies. Although the comparison demonstrates reasonable agreement with the same derived from HMF at intermediate values of $\sigma$ ($\sim 170-250~\mathrm{km}\mathrm{s}^{-1}$), the SDSS catalog underestimates the lens population at both lower and higher $\sigma$ values. At the high end, $\sigma \gtrsim 300\mathrm{km}\mathrm{s}^{-1}$ corresponds to halo masses $\gtrsim 10^{13} M_{\odot}$, which represent group- and cluster-scale haloes rather than individual galaxies; these systems are not included in the SDSS galaxy catalog~\cite{bernardi2003early, blanton2005new}.

At the low end ($\sigma \lesssim 150\,\mathrm{km}\,\mathrm{s}^{-1}$), the discrepancy arises from two factors. First, observational selection effects in magnitude-limited surveys like SDSS systematically exclude low-mass, low-surface-brightness galaxies~\cite{bernardi2003early, blanton2005new, mitchell2005, Choi_2007, bernardi2010}. Second, not all low-mass dark matter haloes host observable early-type galaxies due to inefficient star formation, feedback processes, and environmental effects~\cite{wechsler2018}. These haloes preferentially form late-type or dwarf galaxies rather than the pressure-supported early-type systems for which velocity dispersions are typically measured~\cite{kormendy2009}. Finally, our prescription to convert halo mass to velocity dispersion is relatively simple and does not account for the full complexities of galaxy formation, which we intend to explore in future work.

In Fig.~\ref{fig:optd_hmf_sdss}, we plot the optical depth for the two different lens population models described above ($\sigma$ distribution from SDSS catalog as well as the same derived from the HMF model). Since the HMF is parameterized in terms of cosmological parameters, we also plot the optical depth for different cosmological parameters (same set as Fig.~\ref{fig:hmf_var_with_cosmo}). Since, the HMF-based prescription models a wide range of halo masses (while the SDSS models only galaxy lenses), the optical depth is significantly higher. Interestingly, using a fudge factor of $6.3$ (motivated by~\cite{haris2018}) in the modified Schehter function calculation makes the optical depths comparable for both the prescriptions. 

\section{Summary of anticipated precision}
\label{app:precision_summary}

Figures~\ref{fig:all_combined_2D_posteriors} and \ref{fig:3D_posteriors} show the anticipated posteriors on cosmological parameters derived from a  particular realization of the GW observations, which will suffer from Poisson fluctuations. Tables~\ref{tab:precision_Dominik} and \ref{tab:3d_errors_voyd_and_XG} summaries the anticipated precision after marginalizing over the Poisson fluctuations. Specifically, for each observing scenario, we generate $2000$ independent realizations by drawing $N^{\rm tr}$ and $N_{\ell}^{\rm tr}$ from their corresponding Poisson mean, while also sampling the $N_{\ell}^{\rm tr}$ time delays from the expected distribution at our fiducial cosmology. We then perform the complete inference analysis for each realization and report the median 68\% credible interval widths across all realizations. This gives the typical range of precision achievable in different observing runs. 

\begin{table*}[]
	
	\renewcommand{\arraystretch}{1.4}
	\centering
	\footnotesize 
	\begin{tabular}{c|ccc|c|ccc|c|ccc|c|ccc}
		\hline
		\hline
		&
		\multicolumn{3}{c|}{End of O5} &
		\multirow{4}{*}{} &
		\multicolumn{3}{c|}{End of O6} &
		\multirow{4}{*}{} &
		\multicolumn{3}{c|}{End of Voyager} &
		\multirow{4}{*}{} &
		\multicolumn{3}{c}{End of XG} \\ \hline\hline 
		\multicolumn{16}{c}{Dominik (calibrated with GWTC3 median~\cite{ligo2023gwtc})} \\ \hline\hline 
		\cline{1-4} \cline{6-8} \cline{10-12} \cline{14-16} 
		$H_{0}$ &
		\multicolumn{1}{c|}{$67.9^{+21.6}_{-14.1}$} &
		\multicolumn{1}{c|}{} &
		$69.6^{+5.52}_{-4.76}$ &
		&
		\multicolumn{1}{c|}{$69.3^{+7.39}_{-6.91}$} &
		\multicolumn{1}{c|}{} &
		$69.7^{+1.98}_{-1.91}$ &
		&
		\multicolumn{1}{c|}{$69.8^{+3.46}_{-3.35}$} &
		\multicolumn{1}{c|}{} &
		$69.9^{+1.03}_{-1.00}$ &
		&
		\multicolumn{1}{c|}{$70.0^{+1.30}_{-1.26}$} &
		\multicolumn{1}{c|}{} &
		$69.9^{+0.46}_{-0.46}$ \\ \cline{1-4} \cline{6-8} \cline{10-12} \cline{14-16} 
		$\Omega_{m}$ &
		\multicolumn{1}{c|}{$0.324^{+0.145}_{-0.119}$} &
		\multicolumn{1}{c|}{$0.306^{+0.044}_{-0.038}$} &
		&
		&
		\multicolumn{1}{c|}{$0.313^{+0.055}_{-0.044}$} &
		\multicolumn{1}{c|}{$0.308^{+0.017}_{-0.016}$} &
		&
		&
		\multicolumn{1}{c|}{$0.31^{+0.023}_{-0.020}$} &
		\multicolumn{1}{c|}{$0.31^{+0.009}_{-0.008}$} &
		&
		&
		\multicolumn{1}{c|}{$0.309^{+0.007}_{-0.007}$} &
		\multicolumn{1}{c|}{$0.310^{+0.004}_{-0.004}$} &
		\\ \cline{1-4} \cline{6-8} \cline{10-12} \cline{14-16} 
		$\sigma_{8}$ &
		\multicolumn{1}{c|}{} &
		\multicolumn{1}{c|}{$0.84^{+0.146}_{-0.135}$} &
		$0.837^{+0.136}_{-0.120}$ &
		&
		\multicolumn{1}{c|}{} &
		\multicolumn{1}{c|}{$0.815^{+0.056}_{-0.050}$} &
		$0.815^{+0.044}_{-0.038}$ &
		&
		\multicolumn{1}{c|}{} &
		\multicolumn{1}{c|}{$0.809^{+0.024}_{-0.023}$} &
		$0.810^{+0.018}_{-0.017}$ &
		&
		\multicolumn{1}{c|}{} &
		\multicolumn{1}{c|}{$0.809^{+0.008}_{-0.008}$} &
		$0.810^{+0.006}_{-0.006}$ \\ 
		\hline
		\hline
		\multicolumn{16}{c}{MD without delay (calibrated with GWTC3 median~\cite{ligo2023gwtc})} \\ \hline\hline  
		$H_{0}$ &
		\multicolumn{1}{c|}{$69.2^{+24.0}_{-15.1}$} &
		\multicolumn{1}{c|}{} &
		$69.7^{+5.86}_{-5.21}$ &
		&
		\multicolumn{1}{c|}{$68.8^{+9.73}_{-8.91}$} &
		\multicolumn{1}{c|}{} &
		$69.7^{+2.41}_{-2.29}$ &
		&
		\multicolumn{1}{c|}{$69.7^{+4.99}_{-4.79}$} &
		\multicolumn{1}{c|}{} &
		$69.9^{+1.37}_{-1.33}$ &
		&
		\multicolumn{1}{c|}{$69.8^{+2.08}_{-2.04}$} &
		\multicolumn{1}{c|}{} &
		$69.9^{+0.68}_{-0.67}$ \\ \cline{1-4} \cline{6-8} \cline{10-12} \cline{14-16} 
		$\Omega_{m}$ &
		\multicolumn{1}{c|}{$0.317^{+0.157}_{-0.125}$} &
		\multicolumn{1}{c|}{$0.310^{+0.046}_{-0.041}$} &
		&
		&
		\multicolumn{1}{c|}{$0.316^{+0.078}_{-0.058}$} &
		\multicolumn{1}{c|}{$0.308^{+0.021}_{-0.019}$} &
		&
		&
		\multicolumn{1}{c|}{$0.310^{+0.034}_{-0.029}$} &
		\multicolumn{1}{c|}{$0.309^{+0.011}_{-0.011}$} &
		&
		&
		\multicolumn{1}{c|}{$0.310^{+0.012}_{-0.011}$} &
		\multicolumn{1}{c|}{$0.309^{+0.005}_{-0.005}$} &
		\\ \cline{1-4} \cline{6-8} \cline{10-12} \cline{14-16} 
		$\sigma_{8}$ &
		\multicolumn{1}{c|}{} &
		\multicolumn{1}{c|}{$0.843^{+0.156}_{-0.145}$} &
		$0.846^{+0.148}_{-0.134}$ &
		&
		\multicolumn{1}{c|}{} &
		\multicolumn{1}{c|}{$0.819^{+0.076}_{-0.064}$} &
		$0.818^{+0.062}_{-0.051}$ &
		&
		\multicolumn{1}{c|}{} &
		\multicolumn{1}{c|}{$0.811^{+0.035}_{-0.032}$} &
		$0.810^{+0.027}_{-0.025}$ &
		&
		\multicolumn{1}{c|}{} &
		\multicolumn{1}{c|}{$0.810^{+0.013}_{-0.013}$} &
		$0.809^{+0.01}_{-0.009}$ \\ 
		\hline
		\hline 
	\end{tabular}
	\caption{Median parameter recovery and $68\%$ credible intervals ($\pm 1\sigma$) from $2000$ independent mock realizations of strong lensing observations at our fiducial cosmology $\vOmega_{\rm tr}$, assuming \textit{Dominik} (top) and \textit{MD without delay} (bottom) merger rate models. Pairwise constraints are shown for O5, O6, Voyager, and XG, with one cosmological parameter fixed at its fiducial value while the other two are jointly constrained. Each realization involves Poisson sampling of the expected number of lensed events and independent mergers, random draws of time delays from the predicted distribution at $\vOmega_{\rm tr}$, and the complete Bayesian inference as described in the main text.}
	\label{tab:precision_Dominik}
\end{table*}

\begin{table*}[]
	\renewcommand{\arraystretch}{2}
	\small
	\centering
	\begin{tabular}{c|ccc|cl|ccc}
		\hline
		\hline 
		 &
		\multicolumn{3}{c|}{Dominik} &
		\multicolumn{2}{c|}{\multirow{5}{*}{}} &
		\multicolumn{3}{c}{MD without delay} \\ \cline{1-4} \cline{7-9} 
		 &
		\multicolumn{1}{c|}{optimistic} &
		\multicolumn{1}{c|}{median} &
		conservative &
		\multicolumn{2}{c|}{} &
		\multicolumn{1}{c|}{optimistic} &
		\multicolumn{1}{c|}{median} &
		conservative \\ \cline{1-4} \cline{7-9} 
		$H_{0}$ &
		\multicolumn{1}{c|}{$69.53^{+ 4.69}_{-4.3}$} &
		\multicolumn{1}{c|}{$70.15^{+ 5.79}_{-5.22}$} &
		$70.13^{+ 6.8}_{-5.78}$ &
		\multicolumn{2}{c|}{} &
		\multicolumn{1}{c|}{$70.3^{+ 6.9}_{-5.87}$} &
		\multicolumn{1}{c|}{$71.54^{+ 7.89}_{-6.87}$} &
		$71.56^{+ 8.78}_{-7.37}$ \\ \cline{1-4} \cline{7-9} 
		$\Omega_{m}$ &
		\multicolumn{1}{c|}{$0.312^{+0.036}_{-0.034}$} &
		\multicolumn{1}{c|}{$0.307^{+0.043}_{-0.041}$} &
		$0.308^{+0.048}_{-0.047}$ &
		\multicolumn{2}{c|}{} &
		\multicolumn{1}{c|}{$0.308^{+0.049}_{-0.047}$} &
		\multicolumn{1}{c|}{$0.297^{+0.056}_{-0.052}$} &
		$0.294^{+0.06}_{-0.055}$ \\ \cline{1-4} \cline{7-9} 
		$\sigma_{8}$ &
		\multicolumn{1}{c|}{$0.808^{+0.031}_{-0.03}$} &
		\multicolumn{1}{c|}{$0.81^{+0.038}_{-0.036}$} &
		$0.811^{+0.044}_{-0.04}$ &
		\multicolumn{2}{c|}{} &
		\multicolumn{1}{c|}{$0.81^{+0.045}_{-0.041}$} &
		\multicolumn{1}{c|}{$0.82^{+0.051}_{-0.047}$} &
		$0.824^{+0.057}_{-0.051}$ \\ 
		\hline
		\hline 
	\end{tabular}
	\caption{Median parameter recovery and $68\%$ credible intervals ($\pm 1\sigma$) from $200$ independent mock realizations of strong lensing observations by Voyager and XG detectors using the fiducial cosmology $\vOmega_{\rm tr}$. The left (right) block corresponds to the source redshift distribution model by \textit{Dominik} (\textit{MD without delay}). Each model is calibrated with the optimistic, median and conservative estimate of the BBH merger rate at low-redshifts from GWTC-3 data. The expected bounds on one parameter is estimated after marginalizing over all the others.}
	\label{tab:3d_errors_voyd_and_XG}
\end{table*}

\section{Accuracy of the parameter inference}
\label{app:pp_plot}

\begin{figure}[t]
\centering
\includegraphics[height=2.45in]{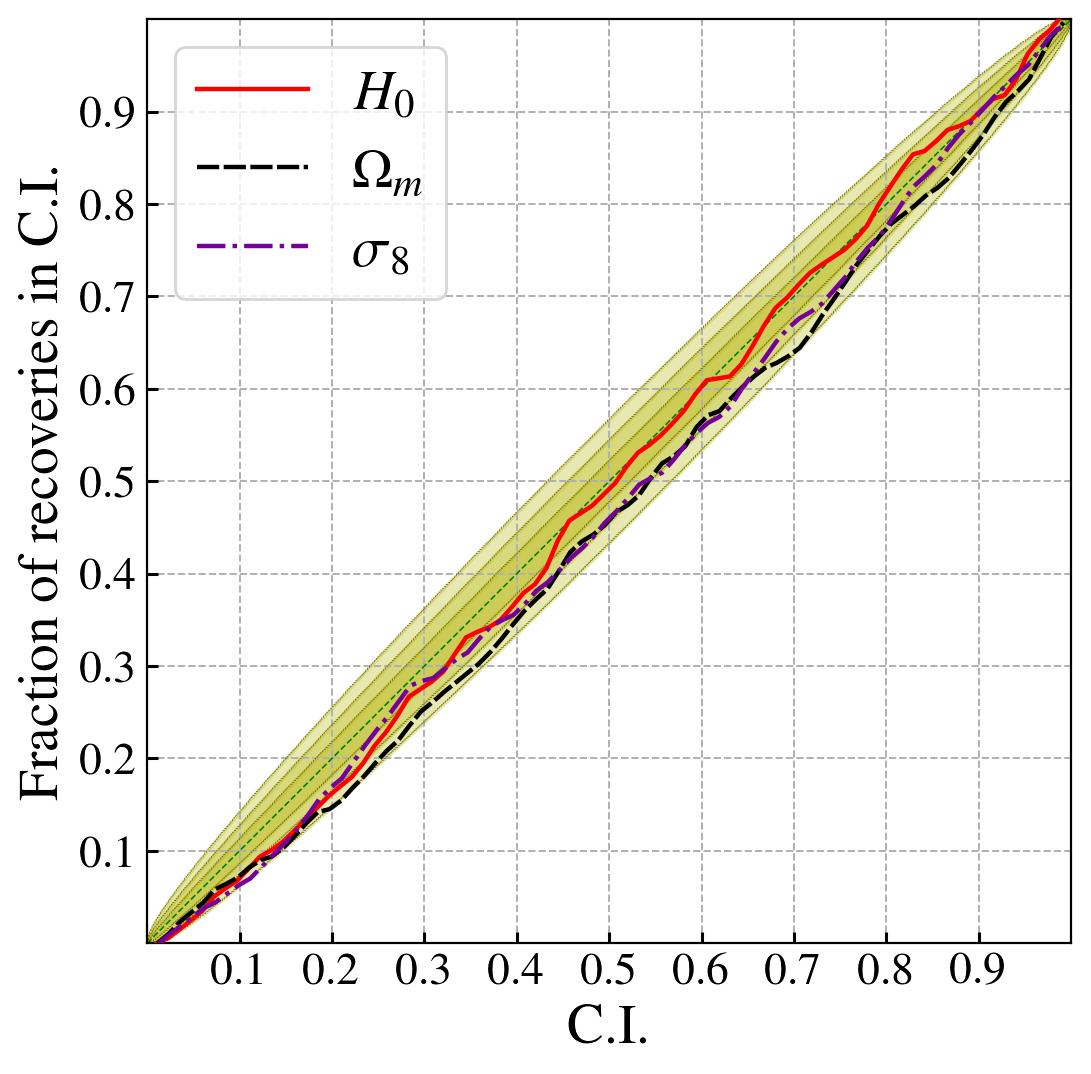}
\caption{Probability-probability ($p-p$) plot assessing accuracy of our inference pipeline (see App.~\ref{app:pp_plot}). The curve shows the cumulative distribution of posterior credible levels evaluated at the true (injected) values. Perfectly calibrated posteriors would follow the diagonal. The shaded regions show the expected statistical fluctuations.} 
\label{fig:pp_plot}
\end{figure}

A Bayesian posterior distribution represents our degree of knowledge about parameter values given the observed data. However, under repeated experiments with different noise realizations, the true parameter values should be distributed within $p$\% credible regions of the posterior for $p\%$ of the realizations. 

To test our analysis pipeline, we employ probability-probability ($p-p$) plots, a frequentest diagnostic tool for evaluating Bayesian posteriors. For each independent realization, we compute the cumulative probability from the marginal posterior distribution of a particular parameter at the injected value. If our inference is unbiased and our uncertainty estimates are accurate, these cumulative probabilities --- drawn from many independent realizations --- should be uniformly distributed between 0 and 1. This should produce a $p-p$ plot that closely follows the diagonal line, with deviations consistent with statistical fluctuations expected from the finite number of realizations. In Fig.~\ref{fig:pp_plot}, we apply this diagnostic to our cosmological inference framework using $500$ independent realizations, resulting in a $p-p$ plot that is diagonal within expected statistical fluctuations.  

\section{Systematic errors due to the imperfect reconstruction of the source redshift distribution}

\begin{figure*}[tbh]
	\centering
	\includegraphics[width=\textwidth]{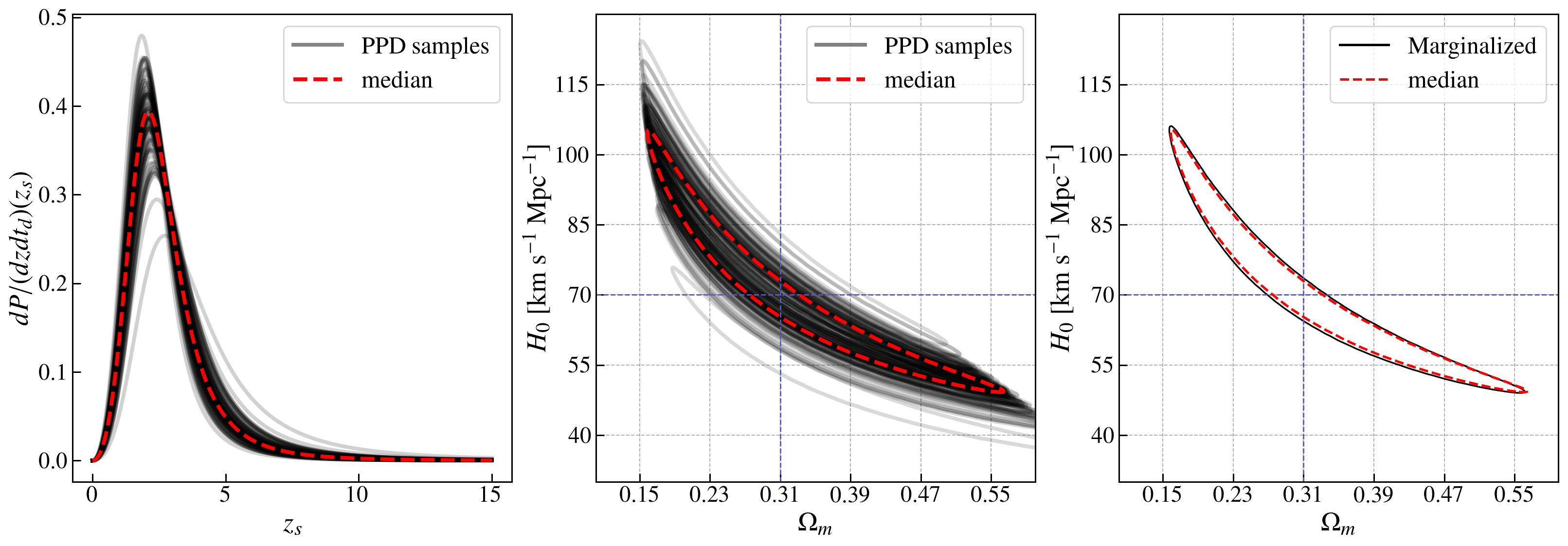}
	\caption{Marginalization over source redshift distribution uncertainties for cosmological inference at the end of O5. \textit{Left:} PPDs of the differential merger rate as a function of source redshift. Gray curves show individual samples drawn from the PPD, while the red dashed curve represents the median distribution used to generate all lensed and unlensed events. \textit{Middle:} Two-dimensional posteriors on $H_0$ and $\Omega_m$ (with $\sigma_8$ fixed at its fiducial value) obtained using individual PPD samples (gray contours) versus the median distribution (red dashed). The spread in gray contours reflects the systematic uncertainty introduced by imperfect knowledge of the source population. \textit{Right:} Fully marginalized posterior (black contour), obtained by marginalizing over all PPD samples, compared to the median-only constraint (red dashed contour). The broad marginalized posterior quantifies the effect of merger rate uncertainties at O5 sensitivity on cosmological constraints.}
	\label{fig:dpdz_marg}
\end{figure*}

The lensing observables (fraction of lensed events as well as time delay distribution) depends also on the redshift distribution of BBH mergers. This can be reconstructed, in principle, by combining luminosity distance posteriors from a large number of (unlensed) BBH mergers under an assumed cosmology~\cite{Vitale_2019, mandel2016, ray2023nonparametric}. Jana et al~\cite{Jana_2024} have shown that these reconstruction errors are negligible for XG detectors. However, from the relatively modest number of detections expected during O5 (see Fig.~\ref{fig:lfraction_and_mrate}), perfect reconstruction of this distribution is not possible. Consequently, we must account for the reconstruction errors by marginalizing over posterior predictive distributions (PPDs).

We construct simplified PPDs for the O5 observing run, as illustrated in Fig.~\ref{fig:dpdz_marg}. Our model assumes the BBH population follows the \emph{MD without delay} model. Parameter uncertainties are estimated by scaling the constraints from the GWTC-3 merger rate analysis. While this approach is conservative --- more sophisticated techniques such as hierarchical parameter estimation would likely yield tighter constraints --- it provides a reasonable assessment of the redshift distribution uncertainty expected at O5 sensitivity.

For our analysis, we define the fiducial redshift distribution as the median of these proxy PPDs (red dotted line in Fig.~\ref{fig:dpdz_marg}, left panel) and generate all lensed events from this median distribution. We then perform cosmological inference in two scenarios: first using individual PPD samples alongside the median (middle panel, Fig.~\ref{fig:dpdz_marg}), and second by fully marginalizing over the PPD ensemble. Comparing these marginalized constraints against those obtained using only the median distribution quantifies the effect in cosmological parameter recovery due to merger rate uncertainties. It can be seen that both the median and marginalized posteriors correctly recover the true parameters.

\newpage
\bibliographystyle{apsrev4-2}
\bibliography{references}

@misc{Barsode:2025agk,
    author = "Barsode, A. and Maity, K. N. and Ajith, P.",
    title = "{Lensing, not luck! Detection prospects of strongly lensed gravitational waves}",
    eprint = "{2510.23238}",
    archivePrefix = "{arXiv}",
    primaryClass = "{gr-qc}",
    month = "10",
    year = "2025"
}

@article{Chen:2024gdn,
    author = "Chen, Hsin-Yu and Ezquiaga, Jose Mar{\'\i}a and Gupta, Ish",
    title = "{Cosmography with next-generation gravitational wave detectors}",
    eprint = "2402.03120",
    archivePrefix = "arXiv",
    primaryClass = "gr-qc",
    doi = "10.1088/1361-6382/ad424f",
    journal = "Class. Quant. Grav.",
    volume = "41",
    number = "12",
    pages = "125004",
    year = "2024"
}

@article{sigma8_tension_2025,
	title = {Exploring the baryonic effect signature in the Hyper Suprime-Cam Year 3 cosmic shear two-point correlations on small scales: The ${S}_{8}$ tension remains present},
	author = {Terasawa, Ryo and others},
	journal = {Phys. Rev. D},
	volume = {111},
	issue = {6},
	pages = {063509},
	numpages = {28},
	year = {2025},
	month = {Mar},
	publisher = {American Physical Society},
	doi = {10.1103/PhysRevD.111.063509},
	url = {https://link.aps.org/doi/10.1103/PhysRevD.111.063509}
}

@article{aasi2015advanced,
	title={Advanced ligo},
	author={Aasi, Junaid and Abbott, BP and Abbott, Richard and Abbott, Thomas and Abernathy, MR and Ackley, Kendall and Adams, Carl and Adams, Thomas and Addesso, Paolo and Adhikari, RX and others},
	journal={Classical and quantum gravity},
	volume={32},
	number={7},
	pages={074001},
	year={2015},
	publisher={IOP Publishing}
}

@article{acernese2014advanced,
	title={Advanced Virgo: a second-generation interferometric gravitational wave detector},
	author={Acernese, Fausto and Agathos, M and Agatsuma, K and Aisa, Damiano and Allemandou, N and Allocca, Aea and Amarni, J and Astone, Pia and Balestri, G and Ballardin, G and others},
	journal={Classical and Quantum Gravity},
	volume={32},
	number={2},
	pages={024001},
	year={2014},
	publisher={IOP Publishing}
}

@article{abbott2019gwtc,
	title={GWTC-1: a gravitational-wave transient catalog of compact binary mergers observed by LIGO and Virgo during the first and second observing runs},
	author={Abbott, Benjamin P and Abbott, Richard and Abbott, TDea and Abraham, S and Acernese, F and Ackley, K and Adams, C and Adhikari, RX and Adya, VB and Affeldt, Christoph and others},
	journal={Physical Review X},
	volume={9},
	number={3},
	pages={031040},
	year={2019},
	publisher={APS}
}

@article{abbott2021gwtc,
	title={GWTC-2: compact binary coalescences observed by LIGO and Virgo during the first half of the third observing run},
	author={Abbott, Richard and Abbott, TD and Abraham, S and Acernese, F and Ackley, K and Adams, A and Adams, C and Adhikari, RX and Adya, VB and Affeldt, Christoph and others},
	journal={Physical Review X},
	volume={11},
	number={2},
	pages={021053},
	year={2021},
	publisher={APS}
}

@article{abbott2024gwtc,
	title={GWTC-2.1: Deep extended catalog of compact binary coalescences observed by LIGO and Virgo during the first half of the third observing run},
	author={Abbott, R and Abbott, TD and Acernese, F and Ackley, K and Adams, C and Adhikari, N and Adhikari, RX and Adya, VB and Affeldt, C and Agarwal, D and others},
	journal={Physical Review D},
	volume={109},
	number={2},
	pages={022001},
	year={2024},
	publisher={APS}
}

@article{ligo2023gwtc,
	title={GWTC-3: compact binary coalescences observed by LIGO and Virgo during the second part of the third observing run},
	author={LIGO Scientific Collaboration and Virgo Collaboration and Kagra Collaboration and others},
	journal={Physical Review X},
	volume={13},
	number={4},
	pages={041039},
	year={2023},
	publisher={American Physical Society}
}

@article{abac2025gwtc,
	title={GWTC-4.0: An Introduction to Version 4.0 of the Gravitational-Wave Transient Catalog},
	author={Abac, AG and Abouelfettouh, I and Acernese, F and Ackley, K and Adhicary, S and Adhikari, D and Adhikari, N and Adhikari, RX and Adkins, VK and Afroz, S and others},
	journal={arXiv preprint arXiv:2508.18080},
	year={2025}
}

@article{venumadhav2020new,
	title={New binary black hole mergers in the second observing run of Advanced LIGO and Advanced Virgo},
	author={Venumadhav, Tejaswi and Zackay, Barak and Roulet, Javier and Dai, Liang and Zaldarriaga, Matias},
	journal={Physical Review D},
	volume={101},
	number={8},
	pages={083030},
	year={2020},
	publisher={APS}
}

@article{zackay2021detecting,
	title={Detecting gravitational waves with disparate detector responses: Two new binary black hole mergers},
	author={Zackay, Barak and Dai, Liang and Venumadhav, Tejaswi and Roulet, Javier and Zaldarriaga, Matias},
	journal={Physical Review D},
	volume={104},
	number={6},
	pages={063030},
	year={2021},
	publisher={APS}
}

@article{olsen2022new,
	title={New binary black hole mergers in the LIGO-Virgo O3a data},
	author={Olsen, Seth and Venumadhav, Tejaswi and Mushkin, Jonathan and Roulet, Javier and Zackay, Barak and Zaldarriaga, Matias},
	journal={Physical Review D},
	volume={106},
	number={4},
	pages={043009},
	year={2022},
	publisher={APS}
}

@article{mehta2025new,
	title={New binary black hole mergers in the LIGO-Virgo O3b data},
	author={Mehta, Ajit Kumar and Olsen, Seth and Wadekar, Digvijay and Roulet, Javier and Venumadhav, Tejaswi and Mushkin, Jonathan and Zackay, Barak and Zaldarriaga, Matias},
	journal={Physical Review D},
	volume={111},
	number={2},
	pages={024049},
	year={2025},
	publisher={APS}
}

@article{wadekar2023new,
	title={New black hole mergers in the LIGO-Virgo O3 data from a gravitational wave search including higher-order harmonics},
	author={Wadekar, Digvijay and Roulet, Javier and Venumadhav, Tejaswi and Mehta, Ajit Kumar and Zackay, Barak and Mushkin, Jonathan and Olsen, Seth and Zaldarriaga, Matias},
	journal={arXiv preprint arXiv:2312.06631},
	year={2023}
}

@article{nitz20191,
	title={1-OGC: The first open gravitational-wave catalog of binary mergers from analysis of public Advanced LIGO data},
	author={Nitz, Alexander H and Capano, Collin and Nielsen, Alex B and Reyes, Steven and White, Rebecca and Brown, Duncan A and Krishnan, Badri},
	journal={The Astrophysical Journal},
	volume={872},
	number={2},
	pages={195},
	year={2019},
	publisher={IOP Publishing}
}

@article{nitz20202,
	title={2-OGC: Open Gravitational-wave Catalog of binary mergers from analysis of public Advanced LIGO and Virgo data},
	author={Nitz, Alexander H and Dent, Thomas and Davies, Gareth S and Kumar, Sumit and Capano, Collin D and Harry, Ian and Mozzon, Simone and Nuttall, Laura and Lundgren, Andrew and T{\'a}pai, M{\'a}rton},
	journal={The Astrophysical Journal},
	volume={891},
	number={2},
	pages={123},
	year={2020},
	publisher={American Astronomical Society}
}

@article{nitz20213,
	title={3-OGC: Catalog of gravitational waves from compact-binary mergers},
	author={Nitz, Alexander H and Capano, Collin D and Kumar, Sumit and Wang, Yi-Fan and Kastha, Shilpa and Sch{\"a}fer, Marlin and Dhurkunde, Rahul and Cabero, Miriam},
	journal={The Astrophysical Journal},
	volume={922},
	number={1},
	pages={76},
	year={2021},
	publisher={IOP Publishing}
}

@article{nitz20234,
	title={4-OGC: Catalog of gravitational waves from compact binary mergers},
	author={Nitz, Alexander H and Kumar, Sumit and Wang, Yi-Fan and Kastha, Shilpa and Wu, Shichao and Sch{\"a}fer, Marlin and Dhurkunde, Rahul and Capano, Collin D},
	journal={The Astrophysical Journal},
	volume={946},
	number={2},
	pages={59},
	year={2023},
	publisher={IOP Publishing}
}

@article{koloniari2025new,
	title={New gravitational wave discoveries enabled by machine learning},
	author={Koloniari, Alexandra E and Koursoumpa, Evdokia C and Nousi, Paraskevi and Lampropoulos, Paraskevas and Passalis, Nikolaos and Tefas, Anastasios and Stergioulas, Nikolaos},
	journal={Machine Learning: Science and Technology},
	volume={6},
	number={1},
	pages={015054},
	year={2025},
	publisher={IOP Publishing}
}

@ARTICLE{ligo_india_2013a,
	author = {{Unnikrishnan}, C.~S.},
	title = "{IndIGO and Ligo-India Scope and Plans for Gravitational Wave Research and Precision Metrology in India}",
	journal = {International Journal of Modern Physics D},
	year = {2013},
	month = jan,
	volume = {22},
	number = {1},
	eid = {1341010},
	pages = {1341010},
	doi = {10.1142/S0218271813410101},
	archivePrefix = {arXiv},
	eprint = {1510.06059}
}

@ARTICLE{ligo_india_2022a,
	author = {{Unnikrishnan}, C.~S.},
	title = "{LIGO-India: A decadal assessment on its scope, relevance, progress and future}",
	journal = {International Journal of Modern Physics D},
	year = 2024,
	month = apr,
	volume = {33},
	eid = {2450025},
	pages = {2450025}
}

@misc{efstathiou2024,
	title={Challenges to the Lambda CDM Cosmology}, 
	author={George Efstathiou},
	year={2024},
	eprint={2406.12106},
	archivePrefix={arXiv},
	primaryClass={astro-ph.CO},
	url={https://arxiv.org/abs/2406.12106}, 
}

@article{peebles2025status,
	title={Status of the $\Lambda$CDM theory: supporting evidence and anomalies},
	author={Peebles, Phillip James E},
	journal={Philosophical Transactions A},
	volume={383},
	number={2290},
	pages={20240021},
	year={2025},
	publisher={The Royal Society}
}

@article{Di_Valentino_2025,
	title={The CosmoVerse White Paper: Addressing observational tensions in cosmology with systematics and fundamental physics},
	volume={49},
	ISSN={2212-6864},
	url={http://dx.doi.org/10.1016/j.dark.2025.101965},
	DOI={10.1016/j.dark.2025.101965},
	journal={Physics of the Dark Universe},
	publisher={Elsevier BV},
	author={Di Valentino and others},
	year={2025},
	month=sep, pages={101965} }

@article{abbott2020prospects,
	title={Prospects for observing and localizing gravitational-wave transients with Advanced LIGO, Advanced Virgo and KAGRA},
	author={Abbott, Benjamin P and Abbott, R and Abbott, TD and Abraham, S and Acernese, F and Ackley, K and Adams, C and Adya, VB and Affeldt, C and Agathos, M and others},
	journal={Living Reviews in Relativity},
	volume={23},
	number={3},
	year={2020}
}

@article{ng2018precise,
	title={Precise LIGO lensing rate predictions for binary black holes},
	author={Ng, Ken KY and Wong, Kaze WK and Broadhurst, Tom and Li, Tjonnie GF},
	journal={Physical Review D},
	volume={97},
	number={2},
	pages={023012},
	year={2018},
	publisher={APS}
}

@article{oguri2018effect,
	title={Effect of gravitational lensing on the distribution of gravitational waves from distant binary black hole mergers},
	author={Oguri, Masamune},
	journal={Monthly Notices of the Royal Astronomical Society},
	volume={480},
	number={3},
	pages={3842--3855},
	year={2018},
	publisher={Oxford University Press}
}

@article{li2018gravitational,
	title={Gravitational lensing of gravitational waves: A statistical perspective},
	author={Li, Shun-Sheng and Mao, Shude and Zhao, Yuetong and Lu, Youjun},
	journal={Monthly Notices of the Royal Astronomical Society},
	volume={476},
	number={2},
	pages={2220--2229},
	year={2018},
	publisher={Oxford University Press}
}

@article{smith2017strong,
	title={Strong-lensing of gravitational waves by galaxy clusters},
	author={Smith, Graham P and Berry, Christopher and Bianconi, Matteo and Farr, Will M and Jauzac, Mathilde and Massey, Richard and Richard, Johan and Robertson, Andrew and Sharon, Keren and Vecchio, Alberto and others},
	journal={Proceedings of the International Astronomical Union},
	volume={13},
	number={S338},
	pages={98--102},
	year={2017},
	publisher={Cambridge University Press}
}

@article{dai2017effect,
	title={Effect of lensing magnification on the apparent distribution of black hole mergers},
	author={Dai, Liang and Venumadhav, Tejaswi and Sigurdson, Kris},
	journal={Physical Review D},
	volume={95},
	number={4},
	pages={044011},
	year={2017},
	publisher={APS}
}

@article{mukherjee2021impact,
	title={Impact of astrophysical binary coalescence time-scales on the rate of lensed gravitational wave events},
	author={Mukherjee, Suvodip and Broadhurst, Tom and Diego, Jose M and Silk, Joseph and Smoot, George F},
	journal={Monthly Notices of the Royal Astronomical Society},
	volume={506},
	number={3},
	pages={3751--3759},
	year={2021},
	publisher={Oxford University Press}
}

@misc{Adhikari:2019zpy,
    author = "Adhikari, Rana X. and others",
    title = "{Astrophysical science metrics for next-generation gravitational-wave detectors}",
    eprint = "1905.02842",
    archivePrefix = "arXiv",
    primaryClass = "astro-ph.HE",
    reportNumber = "LIGO-T1200099",
    doi = "10.1088/1361-6382/ab3cff",
    journal = "Class. Quant. Grav.",
    volume = "36",
    number = "24",
    pages = "245010",
    year = "2019"
}

@article{Jana_2023,
  title = {Cosmography using Strongly Lensed Gravitational Waves from Binary Black Holes},
  author = {Jana, Souvik and Kapadia, Shasvath J. and Venumadhav, Tejaswi and Ajith, Parameswaran},
  journal = {Phys. Rev. Lett.},
  volume = {130},
  number = {26},
  pages = {261401},
  numpages = {8},
  year = {2023},
  month = {Jun},
  publisher = {American Physical Society},
  doi = {10.1103/PhysRevLett.130.261401},
  url = {https://link.aps.org/doi/10.1103/PhysRevLett.130.261401}
}

@misc{Jana_2024,
	author={Jana, Souvik and Kapadia, Shasvath J and Venumadhav, Tejaswi and More, Surhud and Ajith, Parameswaran},
	title={Strong-lensing cosmography using third-generation gravitational-wave detectors},
	journal={Classical and Quantum Gravity},
	url={http://iopscience.iop.org/article/10.1088/1361-6382/ad8d2e},
	year={2024}
}

@misc{haris2018,
	title={Identifying strongly lensed gravitational wave signals from binary black hole mergers}, 
	author={K. Haris and Ajit Kumar Mehta and Sumit Kumar and Tejaswi Venumadhav and Parameswaran Ajith},
	year={2018},
	eprint={1807.07062},
	archivePrefix={arXiv},
	primaryClass={gr-qc},
	url={https://arxiv.org/abs/1807.07062}, 
}

@book{1992schneider,
	author = {{Schneider}, Peter and {Ehlers}, J{\"u}rgen and {Falco}, Emilio E.},
	title = {Gravitational Lenses},
	year = {1992},
	doi = {10.1007/978-3-662-03758-4},
	adsurl = {https://ui.adsabs.harvard.edu/abs/1992grle.book.....S},
	publisher = {Springer Berlin, Heidelberg}
}

@book{creighton2012,
	title={Gravitational-wave physics and astronomy: An introduction to theory, experiment and data analysis},
	author={Creighton, Jolien DE and Anderson, Warren G},
	year={2012},
	publisher={John Wiley \& Sons}
}

@misc{Choi_2007,
	title={Internal and Collective Properties of Galaxies in the Sloan Digital Sky Survey},
	volume={658},
	ISSN={1538-4357},
	url={http://dx.doi.org/10.1086/511060},
	DOI={10.1086/511060},
	number={2},
	journal={The Astrophysical Journal},
	publisher={American Astronomical Society},
	author={Choi, Yun‐Young and Park, Changbom and Vogeley, Michael S.},
	year={2007},
	month=apr, pages={884–897} }

@misc{Behroozi_2013,
	doi = {10.1088/0004-637X/770/1/57},
	url = {https://dx.doi.org/10.1088/0004-637X/770/1/57},
	year = {2013},
	month = {may},
	publisher = {The American Astronomical Society},
	volume = {770},
	number = {1},
	pages = {57},
	author = {Behroozi, Peter S. and Wechsler, Risa H. and Conroy, Charlie},
	title = {THE AVERAGE STAR FORMATION HISTORIES OF GALAXIES IN DARK MATTER HALOS FROM z = 0–8},
	journal = {The Astrophysical Journal}
}

@misc{Dominik_2013,
	title={DOUBLE COMPACT OBJECTS. II. COSMOLOGICAL MERGER RATES},
	volume={779},
	ISSN={1538-4357},
	url={http://dx.doi.org/10.1088/0004-637X/779/1/72},
	DOI={10.1088/0004-637x/779/1/72},
	number={1},
	journal={The Astrophysical Journal},
	publisher={American Astronomical Society},
	author={Dominik, Michal and Belczynski, Krzysztof and Fryer, Christopher and Holz, Daniel E. and Berti, Emanuele and Bulik, Tomasz and Mandel, Ilya and O’Shaughnessy, Richard},
	year={2013},
	month=nov, pages={72} 
}

@misc{Vitale_2019,
	title={Measuring the Star Formation Rate with Gravitational Waves from Binary Black Holes},
	volume={886},
	ISSN={2041-8213},
	url={http://dx.doi.org/10.3847/2041-8213/ab50c0},
	DOI={10.3847/2041-8213/ab50c0},
	number={1},
	journal={The Astrophysical Journal Letters},
	publisher={American Astronomical Society},
	author={Vitale, Salvatore and Farr, Will M. and Ng, Ken K. Y. and Rodriguez, Carl L.},
	year={2019},
	month=nov, pages={L1} }

@misc{Madau_2014,
	title={Cosmic Star-Formation History},
	volume={52},
	ISSN={1545-4282},
	url={http://dx.doi.org/10.1146/annurev-astro-081811-125615},
	DOI={10.1146/annurev-astro-081811-125615},
	number={1},
	journal={Annual Review of Astronomy and Astrophysics},
	publisher={Annual Reviews},
	author={Madau, Piero and Dickinson, Mark},
	year={2014},
	month=aug, pages={415–486} }

@misc{Strolger_2004,
	author = {{Strolger}, Louis-Gregory and others},
	title = "{The Hubble Higher z Supernova Search: Supernovae to z \raisebox{-0.5ex}\textasciitilde 1.6 and Constraints on Type Ia Progenitor Models}",
	journal = {\apj},
	year = 2004,
	month = sep,
	volume = {613},
	number = {1},
	pages = {200-223},
	doi = {10.1086/422901},
	archivePrefix = {arXiv},
	eprint = {astro-ph/0406546}
}

@misc{Belczynski_2008,
	doi = {10.1086/521026},
	url = {https://dx.doi.org/10.1086/521026},
	year = {2008},
	month = {jan},
	publisher = {},
	volume = {174},
	number = {1},
	pages = {223},
	author = {Belczynski, Krzysztof and others},
	title = {Compact Object Modeling with the StarTrack Population Synthesis Code},
	journal = {The Astrophysical Journal Supplement Series}
}

@misc{curtler_1993,
	title = {Gravitational waves from merging compact binaries: How accurately can one extract the binary's parameters from the inspiral waveform?},
	author = {Cutler, Curt and Flanagan, \'Eanna E.},
	journal = {Phys. Rev. D},
	volume = {49},
	issue = {6},
	pages = {2658--2697},
	numpages = {0},
	year = {1994},
	month = {Mar},
	publisher = {American Physical Society},
	doi = {10.1103/PhysRevD.49.2658},
	url = {https://link.aps.org/doi/10.1103/PhysRevD.49.2658}
}

@misc{curtler_1994,
	title = {Gravitational radiation reaction for bound motion around a Schwarzschild black hole},
	author = {Cutler, Curt and Kennefick, Daniel and Poisson, Eric},
	journal = {Phys. Rev. D},
	volume = {50},
	issue = {6},
	pages = {3816--3835},
	numpages = {0},
	year = {1994},
	month = {Sep},
	publisher = {American Physical Society},
	doi = {10.1103/PhysRevD.50.3816},
	url = {https://link.aps.org/doi/10.1103/PhysRevD.50.3816}
}

@article{hogg1999distance,
	title={Distance measures in cosmology},
	author={Hogg, David W},
	journal={arXiv preprint astro-ph/9905116},
	year={1999}
}

@misc{Malmquist_1922,
	author = {{Malmquist}, K.~G.},
	title = "{On some relations in stellar statistics}",
	journal = {Meddelanden fran Lunds Astronomiska Observatorium Serie I},
	year = 1922,
	month = mar,
	volume = {100},
	pages = {1-52},
	adsurl = {https://ui.adsabs.harvard.edu/abs/1922MeLuF.100....1M}
}

@misc{Malmquist_1925,
	author = {{Malmquist}, K.~G.},
	title = "{A contribution to the problem of determining the distribution in space of the stars}",
	journal = {Meddelanden fran Lunds Astronomiska Observatorium Serie I},
	year = 1925,
	month = feb,
	volume = {106},
	pages = {1-12},
	adsurl = {https://ui.adsabs.harvard.edu/abs/1925MeLuF.106....1M}
}

@article{planck18_A&A,
	title={Planck 2018 results-I. Overview and the cosmological legacy of Planck},
	author={Aghanim, Nabila and Akrami, Yashar and Arroja, Frederico and Ashdown, Mark and Aumont, J and Baccigalupi, Carlo and Ballardini, M and Banday, Anthony J and Barreiro, RB and Bartolo, Nicola and others},
	journal={Astronomy \& Astrophysics},
	volume={641},
	pages={A1},
	year={2020},
	publisher={EDP sciences}
}

@article{Riess_2022,
	title={A Comprehensive Measurement of the Local Value of the Hubble Constant with 1 km s−1 Mpc−1 Uncertainty from the Hubble Space Telescope and the SH0ES Team},
	volume={934},
	ISSN={2041-8213},
	url={http://dx.doi.org/10.3847/2041-8213/ac5c5b},
	DOI={10.3847/2041-8213/ac5c5b},
	number={1},
	journal={The Astrophysical Journal Letters},
	publisher={American Astronomical Society},
	author={Riess, Adam G. and Yuan, Wenlong and Macri, Lucas M. and Scolnic, Dan and Brout, Dillon and Casertano, Stefano and Jones, David O. and Murakami, Yukei and Anand, Gagandeep S. and Breuval, Louise and Brink, Thomas G. and Filippenko, Alexei V. and Hoffmann, Samantha and Jha, Saurabh W. and D’arcy Kenworthy, W. and Mackenty, John and Stahl, Benjamin E. and Zheng, WeiKang},
	year={2022},
	month=jul, pages={L7} }

@article{bryan1998,
	title={Statistical properties of x-ray clusters: Analytic and numerical comparisons},
	author={Bryan, Greg L and Norman, Michael L},
	journal={The Astrophysical Journal},
	volume={495},
	number={1},
	pages={80},
	year={1998},
	publisher={IOP Publishing}
}

@article{bernardi2003early,
	title={Early-type galaxies in the Sloan digital sky survey. I. The sample},
	author={Bernardi, Mariangela and Sheth, Ravi K and Annis, James and Burles, Scott and Eisenstein, Daniel J and Finkbeiner, Douglas P and Hogg, David W and Lupton, Robert H and Schlegel, David J and SubbaRao, Mark and others},
	journal={The Astronomical Journal},
	volume={125},
	number={4},
	pages={1817},
	year={2003},
	publisher={IOP Publishing}
}

@article{blanton2005new,
	title={New York University Value-Added Galaxy Catalog: a galaxy catalog based on new public surveys},
	author={Blanton, Michael R and Schlegel, David J and Strauss, Michael A and Brinkmann, J and Finkbeiner, Douglas and Fukugita, Masataka and Gunn, James E and Hogg, David W and Ivezi{\'c}, {\v{Z}}eljko and Knapp, GR and others},
	journal={The Astronomical Journal},
	volume={129},
	number={6},
	pages={2562},
	year={2005},
	publisher={IOP Publishing}
}

@article{mitchell2005,
	title={Improved cosmological constraints from gravitational lens statistics},
	author={Mitchell, Jonathan L and Keeton, Charles R and Frieman, Joshua A and Sheth, Ravi K},
	journal={The Astrophysical Journal},
	volume={622},
	number={1},
	pages={81},
	year={2005},
	publisher={IOP Publishing}
}

@article{bernardi2010,
	title={Galaxy luminosities, stellar masses, sizes, velocity dispersions as a function of morphological type},
	author={Bernardi, Mariangela and Shankar, Francesco and Hyde, JB and Mei, Simona and Marulli, Federico and Sheth, Ravi K},
	journal={Monthly Notices of the Royal Astronomical Society},
	volume={404},
	number={4},
	pages={2087--2122},
	year={2010},
	publisher={Blackwell Publishing Ltd Oxford, UK}
}

@article{wechsler2018,
	title={The connection between galaxies and their dark matter halos},
	author={Wechsler, Risa H and Tinker, Jeremy L},
	journal={Annual Review of Astronomy and Astrophysics},
	volume={56},
	number={1},
	pages={435--487},
	year={2018},
	publisher={Annual Reviews}
}

@article{kormendy2009,
	title={Structure and formation of elliptical and spheroidal galaxies},
	author={Kormendy, John and Fisher, David B and Cornell, Mark E and Bender, Ralf},
	journal={The Astrophysical Journal Supplement Series},
	volume={182},
	number={1},
	pages={216},
	year={2009},
	publisher={IOP Publishing}
}

@misc{H1L1V1-psd-O3O4O5,
	author = "{LIGO-Virgo Collaboration}",
	title = "Noise curves used for Simulations in the update of the Observing Scenarios Paper",
	note = "{LIGO} Document T2000012-v1",
	url = "https://dcc.ligo.org/LIGO-T2000012/public",
	year ="2020"
}

@misc{HLA-psd-O6,
	author       = {{LIGO-Virgo Collaboration}},
	title        = {A{\#} {S}train {S}ensitivity},
	year         = {2023},
	note         = "{LIGO} Document T2300041-v1",
	url          = {https://dcc.ligo.org/LIGO-T2300041/public},
	urldate      = {2025-08-28},
}

@misc{HLV-psd-O4a,
	author       = {{LIGO-Virgo Collaboration}},
	title        = {GWIStat},
	year       = {2023},
	url          = {https://git.ligo.org/computing/services/gwistat/-/tree/master/psd},
	urldate      = {2025-08-28},
}

@misc{HLA-voyager,
	author       = {{LIGO-Virgo Collaboration}},
	title        = {Report from the LSC Post-O5 Study Group},
	year         = {2023},
	note         = {{LIGO} Document T2200287–v2},
	url          = {https://dcc.ligo.org/public/0183/T2200287/002/T2200287v2_PO5report.pdf},
}

@article{maggiore2020science,
	title={Science case for the Einstein telescope},
	author={Maggiore, Michele and Van Den Broeck, Chris and Bartolo, Nicola and Belgacem, Enis and Bertacca, Daniele and Bizouard, Marie Anne and Branchesi, Marica and Clesse, Sebastien and Foffa, Stefano and Garc{\'\i}a-Bellido, Juan and others},
	journal={Journal of Cosmology and Astroparticle Physics},
	volume={2020},
	number={03},
	pages={050},
	year={2020},
	publisher={IOP Publishing}
}

@article{hild2011sensitivity,
	title={Sensitivity studies for third-generation gravitational wave observatories},
	author={Hild, S and Abernathy, M and Acernese, F ea and Amaro-Seoane, P and Andersson, N and Arun, K and Barone, F and Barr, B and Barsuglia, M and Beker, M and others},
	journal={Classical and Quantum gravity},
	volume={28},
	number={9},
	pages={094013},
	year={2011},
	publisher={IOP Publishing}
}

@misc{Borhanian_2021,
	title={GWBENCH: a novel Fisher information package for gravitational-wave benchmarking},
	volume={38},
	ISSN={1361-6382},
	url={http://dx.doi.org/10.1088/1361-6382/ac1618},
	DOI={10.1088/1361-6382/ac1618},
	number={17},
	journal={Classical and Quantum Gravity},
	publisher={IOP Publishing},
	author={Borhanian, S},
	year={2021},
	month=aug, pages={175014}}

@misc{Reitze_2019,
	title = {Cosmic Explorer: The U.S. Contribution to Gravitational-Wave Astronomy beyond LIGO},
	author = "Reitze, David and others",
	eprint = "1907.04833",
	archivePrefix = "arXiv",
	primaryClass = "astro-ph.IM",
	reportNumber = "LIGO-P1900316",
	journal = "Bull. Am. Astron. Soc.",
	volume = "51",
	number = "7",
	pages = "035",
	year = "2019"
}

@misc{evans2021horizonstudycosmicexplorer,
	title={A Horizon Study for Cosmic Explorer: Science, Observatories, and Community}, 
	author={Matthew Evans and others},
	year={2021},
	eprint={2109.09882},
	archivePrefix={arXiv},
	primaryClass={astro-ph.IM},
	url={https://arxiv.org/abs/2109.09882}, 
}

@misc{lalsuite_det_locs,
	author         = "{LIGO Scientific Collaboration} and {Virgo Collaboration} and {KAGRA Collaboration}",
	title          = "{LVK} {A}lgorithm {L}ibrary - {LALS}uite",
	howpublished   = "Free software (GPL)",
	doi            = "10.7935/GT1W-FZ16",
	year           = "2018"
}

@misc{GraceDB_public,
	author         = "{LIGO Scientific Collaboration} and {Virgo Collaboration} and {KAGRA Collaboration}",
	title          = "{LIGO}/{V}irgo/{KAGRA} Public Alerts",
	url = "https://gracedb.ligo.org/superevents/public/O4/",
	year = "2025"
}

@article{mandel2016,
	title={Model-independent inference on compact-binary observations},
	author={Mandel, Ilya and Farr, Will M and Colonna, Andrea and Stevenson, Simon and Ti{\v{n}}o, Peter and Veitch, John},
	journal={Monthly Notices of the Royal Astronomical Society},
	pages={stw2883},
	year={2016},
	publisher={Oxford University Press}
}

@article{ray2023nonparametric,
	title={Nonparametric inference of the population of compact binaries from gravitational-wave observations using binned gaussian processes},
	author={Ray, Anarya and Hernandez, Ignacio Maga{\~n}a and Mohite, Siddharth and Creighton, Jolien and Kapadia, Shasvath},
	journal={The Astrophysical Journal},
	volume={957},
	number={1},
	pages={37},
	year={2023},
	publisher={IOP Publishing}
}

@article{birrer2025tdcosmo,
	title={TDCOSMO 2025: Cosmological constraints from strong lensing time delays},
	author={Birrer, Simon and Buckley-Geer, Elizabeth J and Cappellari, Michele and Courbin, Fr{\'e}d{\'e}ric and Dux, Fr{\'e}d{\'e}ric and Fassnacht, Christopher D and Frieman, Joshua A and Galan, Aymeric and Gilman, Daniel and Huang, Xiang-Yu and others},
	journal={arXiv preprint arXiv:2506.03023},
	year={2025}
}

@article{abbott2017_bright_siren,
	title={A Gravitational-Wave Standard Siren Measurement of the Hubble Constant},
	author={Abbott, BP and LIGO Scientific Collaboration and Virgo Collaboration and others},
	journal={Nature (London)},
	volume={551},
	pages={85},
	year={2017}
}

@article{abbott_2023a,
	author = "Abbott, R. and others",
	collaboration = "LIGO Scientific, Virgo, KAGRA",
	title = "{Constraints on the Cosmic Expansion History from GWTC{\textendash}3}",
	eprint = "2111.03604",
	archivePrefix = "arXiv",
	primaryClass = "astro-ph.CO",
	reportNumber = "LIGO-P2100185-v6, LIGO-P2100185-v5",
	doi = "10.3847/1538-4357/ac74bb",
	journal = "Astrophys. J.",
	volume = "949",
	number = "2",
	pages = "76",
	year = "2023"
}

@article{dalya2018glade,
	title={GLADE: A galaxy catalogue for multimessenger searches in the advanced gravitational-wave detector era},
	author={D{\'a}lya, Gergely and Galg{\'o}czi, G{\'a}bor and Dobos, L{\'a}szl{\'o} and Frei, Zsolt and Heng, Ik Siong and Macas, Ronaldas and Messenger, Christopher and Raffai, P{\'e}ter and de Souza, Rafael S},
	journal={Monthly Notices of the Royal Astronomical Society},
	volume={479},
	number={2},
	pages={2374--2381},
	year={2018},
	publisher={Oxford University Press}
}

@article{dalya2022glade+,
	title={GLADE+: an extended galaxy catalogue for multimessenger searches with advanced gravitational-wave detectors},
	author={D{\'a}lya, Gergely and D{\'\i}az, R and Bouchet, FR and Frei, Z and Jasche, Jens and Lavaux, G and Macas, R and Mukherjee, S and P{\'a}lfi, M and De Souza, RS and others},
	journal={Monthly Notices of the Royal Astronomical Society},
	volume={514},
	number={1},
	pages={1403--1411},
	year={2022},
	publisher={Oxford University Press}
}

@article{palmese2023standard,
	title={A standard siren measurement of the Hubble constant using gravitational-wave events from the first three LIGO/Virgo observing runs and the DESI legacy survey},
	author={Palmese, Antonella and Bom, Clecio R and Mucesh, Sunil and Hartley, William G},
	journal={The Astrophysical Journal},
	volume={943},
	number={1},
	pages={56},
	year={2023},
	publisher={IOP Publishing}
}

@article{mukherjee2024cross,
	title={Cross-correlating Dark Sirens and Galaxies: Constraints on H 0 from GWTC-3 of LIGO--Virgo--KAGRA},
	author={Mukherjee, Suvodip and Krolewski, Alex and Wandelt, Benjamin D and Silk, Joseph},
	journal={The Astrophysical Journal},
	volume={975},
	number={2},
	pages={189},
	year={2024},
	publisher={IOP Publishing}
}

@article{chen2024cosmography,
	title={Cosmography with next-generation gravitational wave detectors},
	author={Chen, Hsin-Yu and Ezquiaga, Jose Mar{\'\i}a and Gupta, Ish},
	journal={Classical and Quantum Gravity},
	volume={41},
	number={12},
	pages={125004},
	year={2024},
	publisher={IOP Publishing}
}

@article{afroz2024prospect,
	title={Prospect of precision cosmology and testing general relativity using binary black holes--galaxies cross-correlation},
	author={Afroz, Samsuzzaman and Mukherjee, Suvodip},
	journal={Monthly Notices of the Royal Astronomical Society},
	volume={534},
	number={2},
	pages={1283--1298},
	year={2024},
	publisher={Oxford University Press}
}

@article{alfradique2025systematic,
	title={Systematic bias in dark siren statistical methods and its impact on Hubble constant measurement},
	author={Alfradique, Viviane and Bom, Cl{\'e}cio R and Castro, Tiago},
	journal={Physical Review D},
	volume={112},
	number={6},
	pages={063561},
	year={2025},
	publisher={APS}
}

@article{muttoni2023dark,
	title={Dark siren cosmology with binary black holes in the era of third-generation gravitational wave detectors},
	author={Muttoni, Niccol{\`o} and Laghi, Danny and Tamanini, Nicola and Marsat, Sylvain and Izquierdo-Villalba, David},
	journal={Physical Review D},
	volume={108},
	number={4},
	pages={043543},
	year={2023},
	publisher={APS}
}

@misc{barsode2025lensing,
	title={Lensing, not luck! Detection prospects of strongly lensed gravitational waves}, 
	author={A. Barsode and K. N. Maity and P. Ajith},
	year={2025},
	eprint={2510.23238},
	archivePrefix={arXiv},
	primaryClass={gr-qc},
	url={https://arxiv.org/abs/2510.23238}, 
}

@article{hannuksela2019search,
	title        = {Search for gravitational lensing signatures in LIGO-Virgo binary black hole events},
	author       = {Hannuksela, O.A. and Haris, K. and Ng, K.K.Y. and Kumar, S. and Mehta, A.K. and Keitel, D. and Li, T.G.F. and Ajith, P.},
	year         = 2019,
	journal      = {Astrophys. J. Lett.},
	volume       = 874,
	number       = 1,
	pages        = {L2},
	doi          = {10.3847/2041-8213/ab0c0f},
	archiveprefix = {arXiv},
	eprint       = {1901.02674},
	primaryclass = {gr-qc},
	reportnumber = {LIGO Document P1800297, LIGO-P1800297}
}

@article{LIGOScientific:2021izm,
	title        = {Search for Lensing Signatures in the Gravitational-Wave Observations from the First Half of LIGO\textendash{}Virgo\textquoteright{}s Third Observing Run},
	author       = {Abbott, R. and others},
	year         = 2021,
	month        = 5,
	journal      = {Astrophys. J.},
	volume       = 923,
	number       = 1,
	pages        = 14,
	doi          = {10.3847/1538-4357/ac23db},
	collaboration = {LIGO Scientific, VIRGO},
	eprint       = {2105.06384},
	archiveprefix = {arXiv},
	primaryclass = {gr-qc},
	reportnumber = {LIGO-P2000400}
}

@article{abbott2023search,
	doi = {10.3847/1538-4357/ad3e83},
	url = {https://dx.doi.org/10.3847/1538-4357/ad3e83},
	year = {2024},
	month = {jul},
	publisher = {The American Astronomical Society},
	volume = {970},
	number = {2},
	pages = {191},
	author = {R. Abbott and H. Abe and F. Acernese and K. Ackley and others},
	title = {Search for Gravitational-lensing Signatures in the Full Third Observing Run of the LIGO–Virgo Network},
	journal = {The Astrophysical Journal}
}

@dataset{ligo_scientific_collaboration_and_virgo_2024_10841987,
	author       = {Abbott, R and Abe, H and Acernese, F and Ackley, K and Adhicary, S and Adhikari, N and Adhikari, RX and Adkins, VK and Adya, VB and Affeldt, C and others},
	title        = {{The data for "Search for gravitational-lensing
	signatures in the full third observing run of the
	LIGO–Virgo network"}},
	month        = mar,
	year         = 2024,
	publisher    = {Zenodo},
	doi          = {10.5281/zenodo.10841987},
	url          = {https://doi.org/10.5281/zenodo.10841987}
}

@article{li2023targeted,
	author = "Li, Alvin K. Y. and Lo, Rico K. L. and Sachdev, Surabhi and Chan, Juno C. L. and Lin, E. T. and Li, Tjonnie G. F. and Weinstein, Alan J.",
	collaboration = "LIGO Scientific, Virgo",
	title = "{Targeted subthreshold search for strongly lensed gravitational-wave events}",
	eprint = "1904.06020",
	archivePrefix = "arXiv",
	primaryClass = "gr-qc",
	doi = "10.1103/PhysRevD.107.123014",
	journal = "Phys. Rev. D",
	volume = "107",
	number = "12",
	pages = "123014",
	year = "2023"
}

@article{mcisaac2020search,
	title        = {{Search for strongly lensed counterpart images of binary black hole mergers in the first two LIGO observing runs}},
	author       = {McIsaac, Connor and Keitel, David and Collett, Thomas and Harry, Ian and Mozzon, Simone and Edy, Oliver and Bacon, David},
	year         = 2020,
	journal      = {Phys. Rev. D},
	volume       = 102,
	number       = 8,
	pages        = {084031},
	doi          = {10.1103/PhysRevD.102.084031},
	eprint       = {1912.05389},
	archiveprefix = {arXiv},
	primaryclass = {gr-qc},
	reportnumber = {LIGO-P1900360}
}

@article{dai2020search,
	title        = {Search for Lensed Gravitational Waves Including Morse Phase Information: An Intriguing Candidate in O2},
	author       = {Dai, Liang and Zackay, Barak and Venumadhav, Tejaswi and Roulet, Javier and Zaldarriaga, Matias},
	year         = 2020,
	month        = 7,
	journal={arXiv preprint arXiv:2007.12709},
	archiveprefix = {arXiv},
	eprint       = {2007.12709},
	primaryclass = {astro-ph.HE}
}

@article{janquart2023follow,
	title={Follow-up analyses to the O3 LIGO--Virgo--KAGRA lensing searches},
	volume={526},
	ISSN={1365-2966},
	url={http://dx.doi.org/10.1093/mnras/stad2909},
	DOI={10.1093/mnras/stad2909},
	number={3},
	journal={Monthly Notices of the Royal Astronomical Society},
	publisher={Oxford University Press (OUP)},
	author={Janquart, J and others},
	year={2023},
	month=sep,
	pages={3832–3860}
}

@article{more2022improved,
	title={Improved statistic to identify strongly lensed gravitational wave events},
	author={More, Anupreeta and More, Surhud},
	journal={Monthly Notices of the Royal Astronomical Society},
	volume={515},
	number={1},
	pages={1044--1051},
	year={2022},
	publisher={Oxford University Press}
}

@article{Holz_2005,
	author = "Holz, Daniel E. and Hughes, Scott A.",
	title = "{Using gravitational-wave standard sirens}",
	eprint = "astro-ph/0504616",
	archivePrefix = "arXiv",
	doi = "10.1086/431341",
	journal = "Astrophys. J.",
	volume = "629",
	pages = "15--22",
	year = "2005"
}

@article{Dalal_2006,
	author = "Dalal, Neal and Holz, Daniel E. and Hughes, Scott A. and Jain, Bhuvnesh",
	title = "{Short grb and binary black hole standard sirens as a probe of dark energy}",
	eprint = "astro-ph/0601275",
	archivePrefix = "arXiv",
	doi = "10.1103/PhysRevD.74.063006",
	journal = "Phys. Rev. D",
	volume = "74",
	pages = "063006",
	year = "2006"
}

@article{Nissanke_2009,
	author = "Nissanke, Samaya and Holz, Daniel E. and Hughes, Scott A. and Dalal, Neal and Sievers, Jonathan L.",
	title = "{Exploring short gamma-ray bursts as gravitational-wave standard sirens}",
	eprint = "0904.1017",
	archivePrefix = "arXiv",
	primaryClass = "astro-ph.CO",
	doi = "10.1088/0004-637X/725/1/496",
	journal = "Astrophys. J.",
	volume = "725",
	pages = "496--514",
	year = "2010"
}

@ARTICLE{2019ApJ_Fishbach,
	author = {{Fishbach}, M. and others},
	title = "{A Standard Siren Measurement of the Hubble Constant from GW170817 without the Electromagnetic Counterpart}",
	journal = {\apjl},
	year = 2019,
	month = jan,
	volume = {871},
	number = {1},
	eid = {L13},
	pages = {L13},
	doi = {10.3847/2041-8213/aaf96e},
	archivePrefix = {arXiv},
	eprint = {1807.05667},
	primaryClass = {astro-ph.CO},
	adsurl = {https://ui.adsabs.harvard.edu/abs/2019ApJ...871L..13F}
}

@ARTICLE{Schutz_1986Nature,
	author = {{Schutz}, B.~F.},
	title = "{Determining the Hubble constant from gravitational wave observations}",
	journal = {\nat},
	year = 1986,
	month = sep,
	volume = {323},
	number = {6086},
	pages = {310-311},
	doi = {10.1038/323310a0},
	adsurl = {https://ui.adsabs.harvard.edu/abs/1986Natur.323..310S}
}

@article{Del_Pozzo_2012,
	title={Inference of cosmological parameters from gravitational waves: Applications to second generation interferometers},
	volume={86},
	ISSN={1550-2368},
	url={http://dx.doi.org/10.1103/PhysRevD.86.043011},
	DOI={10.1103/physrevd.86.043011},
	number={4},
	journal={Physical Review D},
	publisher={American Physical Society (APS)},
	author={Del Pozzo, Walter},
	year={2012},
	month=aug }

@article{Nair_2014,
	title = {Measuring the Hubble constant: Gravitational wave observations meet galaxy clustering},
	author = {Nair, Remya and Bose, Sukanta and Saini, Tarun Deep},
	journal = {Phys. Rev. D},
	volume = {98},
	issue = {2},
	pages = {023502},
	numpages = {14},
	year = {2018},
	month = {Jul},
	publisher = {American Physical Society},
	doi = {10.1103/PhysRevD.98.023502},
	url = {https://link.aps.org/doi/10.1103/PhysRevD.98.023502}
}

@article{Soares_Santos_2019,
	title={First Measurement of the Hubble Constant from a Dark Standard Siren using the Dark Energy Survey Galaxies and the LIGO/Virgo Binary–Black-hole Merger GW170814},
	volume={876},
	ISSN={2041-8213},
	url={http://dx.doi.org/10.3847/2041-8213/ab14f1},
	DOI={10.3847/2041-8213/ab14f1},
	number={1},
	journal={The Astrophysical Journal Letters},
	publisher={American Astronomical Society},
	author={Soares-Santos, M. and others},
	year={2019},
	month=apr, pages={L7} }

@article{gray_2020,
	title = {Cosmological inference using gravitational wave standard sirens: A mock data analysis},
	author = {Gray, Rachel and Hernandez, Ignacio Maga\~na and Qi, Hong and Sur, Ankan and Brady, Patrick R. and Chen, Hsin-Yu and Farr, Will M. and Fishbach, Maya and Gair, Jonathan R. and Ghosh, Archisman and Holz, Daniel E. and Mastrogiovanni, Simone and Messenger, Christopher and Steer, Dani\`ele A. and Veitch, John},
	journal = {Phys. Rev. D},
	volume = {101},
	issue = {12},
	pages = {122001},
	numpages = {22},
	year = {2020},
	month = {Jun},
	publisher = {American Physical Society},
	doi = {10.1103/PhysRevD.101.122001},
	url = {https://link.aps.org/doi/10.1103/PhysRevD.101.122001}
}

@article{Borhanian_2020,
	author = "Borhanian, Ssohrab and Dhani, Arnab and Gupta, Anuradha and Arun, K. G. and Sathyaprakash, B. S.",
	title = "{Dark Sirens to Resolve the Hubble{\textendash}Lema{\^\i}tre Tension}",
	eprint = "2007.02883",
	archivePrefix = "arXiv",
	primaryClass = "astro-ph.CO",
	reportNumber = "LIGO document number LIGO-P2000229",
	doi = "10.3847/2041-8213/abcaf5",
	journal = "Astrophys. J. Lett.",
	volume = "905",
	number = "2",
	pages = "L28",
	year = "2020"
}

@article{ezquiaga2022spectral,
	title={Spectral sirens: Cosmology from the full mass distribution of compact binaries},
	author={Ezquiaga, Jose Mar{\'\i}a and Holz, Daniel E},
	journal={Physical Review Letters},
	volume={129},
	number={6},
	pages={061102},
	year={2022},
	publisher={APS}
}

@article{Mali_2024,
	author = "Mali, Utkarsh and Essick, Reed",
	title = "{Striking a Chord with Spectral Sirens: Multiple Features in the Compact Binary Population Correlate with H$_{0}$}",
	eprint = "2410.07416",
	archivePrefix = "arXiv",
	primaryClass = "astro-ph.HE",
	doi = "10.3847/1538-4357/ad9de7",
	journal = "Astrophys. J.",
	volume = "980",
	number = "1",
	pages = "85",
	year = "2025"
}
\end{document}